\documentclass[10pt]{article}

\usepackage[a4paper, margin=0.8in]{geometry} 

\providecommand{\keywords}[1]{\textbf{\textit{Keywords---}} #1}

\usepackage[utf8]{inputenc} 
\usepackage[T1]{fontenc}    
\usepackage{hyperref} 
\usepackage{url}  
\usepackage{booktabs} 
\usepackage{amsfonts} 
\usepackage{nicefrac} 
\usepackage{microtype} 
\usepackage{graphicx}
\usepackage{cite}
\usepackage{doi}

\usepackage{enumitem}
\setlist[itemize]{label=$\triangleright$}

\usepackage{wrapstuff}

\usepackage{bm}
\usepackage{amsmath}
\usepackage{algorithm}
\usepackage{algpseudocode}

\DeclareMathOperator*{\argmax}{arg\,max}
\usepackage[table]{xcolor}
\usepackage{changepage}
\usepackage{multirow}
\usepackage{bbm}
\usepackage{todonotes}
\usepackage[normalem]{ulem}
\usepackage{setspace}
\usepackage{listings}
\usepackage{authblk}

\title{Deviations from Tradition: Stylized Facts in the Era of DeFi}

\author[1]{Daniele Maria Di Nosse\textsuperscript{*,\dag}}
\author[1]{Federico Gatta\textsuperscript{*,\dag}}
\author[1,2]{Fabrizio Lillo}
\author[3,4]{Sebastian Jaimungal}

\affil[1]{Scuola Normale Superiore, Pisa, Italy}
\affil[2]{Dipartimento di Matematica, Universit\`{a} di Bologna, Italy}
\affil[3]{Department of Statistical Sciences, University of Toronto, Canada}
\affil[4]{Oxford-Man Institute for Quantitative Finance, University of Oxford, United Kingdom}

\begin{document}

\maketitle
\begin{center}
  \footnotesize
  * These authors contributed equally to this work.\\
  \dag Corresponding authors. Email: \texttt{daniele.dinosse@sns.it}, \texttt{federico.gatta@sns.it}
\end{center}

\begin{abstract}
    Decentralized Exchanges (DEXs) are now a significant component of the financial world where billions of dollars are traded daily. Differently from traditional markets, which are typically based on Limit Order Books,
    DEXs typically work as Automated Market Makers, and, since the implementation of Uniswap v3, feature concentrated liquidity. By investigating the twenty-four most active pools in Uniswap v3 during 2023 and 2024,
    we empirically study how this structural change in the organization of the markets modifies the well-studied stylized facts of prices, liquidity, and order flow observed in traditional markets. We find a series of new statistical regularities in the distributions and cross-autocorrelation functions of these variables that we are able to associate either with the market structure (e.g., the execution of orders in blocks) or with the intense activity of Maximal Extractable Value searchers, such as Just-in-Time liquidity providers and sandwich attackers. 
    
\end{abstract}


\section{Introduction}
Financial markets are physical or virtual venues where assets such as stocks, commodities, or currencies are traded. Traditionally, they have been dominated by Centralized Exchanges (CEXs), where a central party safeguards assets and oversees transactions. In recent years, however, Decentralized Exchanges (DEXs) have gained attraction by leveraging advancements in blockchain technology to eliminate the need for intermediaries. In this paper, we study how the trading organization of Decentralized Finance (DeFi) modifies the typical stylized facts and statistical properties of prices, orders, and volume observed in Traditional Finance (TradFi). Indeed, nowadays, billions of dollars are traded daily on these platforms\footnote{See \url{https://defillama.com} for updates in the trading activity across DEXs.}.
DEXs operate on the blockchain using specialized programs called smart contracts \cite{szabo1997formalizing}. In the majority of DEXs, these contracts are implemented as Automated Market Makers (AMMs), which are algorithms that manage liquidity and regulate asset pricing on the exchange \cite{alotaibi2021smart}.

AMMs can be thought of as exchange desks and rely on two main components: liquidity pools and price formation rules. Liquidity pools can be conceptualized as ``digital vaults'' containing two types of coins, also referred to as tokens. In the following, we call them token X and token Y. The volumes of tokens stored in these pools are called reserves. As there is no centralized entity supplying liquidity, AMMs have two types of participants: Liquidity Providers (LPs) and Liquidity Takers (LTs). LPs contribute to liquidity by depositing both tokens into the pool, the amounts of which must satisfy specific constraints. In return, they receive a certain share of the pool, for example, in terms of  Liquidity Pool tokens (LP-tokens) or via Non-Fungible Tokens (NFTs). Conversely, LTs interact with the pool by swapping one token for another and paying a fee for this service. These fees are distributed among LPs, typically in proportion to their pool share. To summarize, there are three admissible basic events in a DEX: swap, mint, and burn. A swap occurs when an LT exchanges one token for the other. A mint represents the action of an LP that provides liquidity to the pool. A burn corresponds to an LP withdrawing liquidity from the pool. Finally, it is worth underlining that every operation in the blockchain, independently of its nature, is subject to network-level execution costs paid to the blockchain, such as the gas fee in the Ethereum chain or the transaction fee in Solana. The value of such fees depends on the network congestion, and it is typically made up of two components: i) the base fee -- which is the minimum amount to be paid, and ii) the priority fee -- which is an incentive paid to the network, chosen by the user, for being executed faster than others.

The price formation in AMMs is governed by fixed mathematical rules. The most widespread AMM mechanism is the Constant Function Market Maker, in which the value of a function of the reserves in the pool must be constant before and after every swap. Among all the possible functions, the most widely adopted is the Constant Product Rule, and such AMMs are termed Constant Product Market Makers (CPMMs). Under this rule, the product of the reserves of the two tokens must remain constant before and after a swap. This ensures that the price of one token relative to the other adjusts dynamically based on the reserves' balances. Finally, a mint (burn) shifts the constant product curve upwards (downwards) while keeping the price fixed. A prominent example of a DEX is Uniswap. Among the various blockchains on which Uniswap is deployed, the Ethereum implementation consistently accounts for the highest trading volume. When considering all DEXs and blockchain ecosystems, PancakeSwap \cite{coinmarketcap_dex_rankings} -- a fork of Uniswap deployed on the Binance Smart Chain (BSC) -- currently ranks first by traded volume. Despite this, our analysis focuses on the Ethereum blockchain, motivated by the vast number of projects developed on it, the presence of a dynamic and innovative community, and its comparatively higher degree of decentralization and censorship resistance relative to BSC.

 At the time of writing, there are three operating versions of Uniswap: v2 \cite{adams2020uniswap}, v3 \cite{adams2021uniswap}, and v4 \cite{UniswapV4Whitepaper}. Historically, v1 (no longer working) was the first proposal, while v2 introduced significant improvements. Later, v3 brought important innovations that allowed for Concentrated Liquidity, and it is by far (at the time of writing) the most traded Uniswap version. Finally, v4 introduced even more possibilities for the users with the so-called `hooks' that allow for a greater refinement of the dynamics of the liquidity pools. Although Uniswap v4 is the most complex version, both practically and theoretically, our analysis focuses on Uniswap v3 because Uniswap v4 is only recently released; hence, data is limited and not sufficient for studying the statistical properties of the pools' dynamics. Nonetheless, as the skeleton of Uniswap v3 and v4 is roughly the same, we expect our results to hold across both.

Since DEXs appeared, the scientific community's interest has consistently grown. Several papers, mostly addressing the problem of optimizing liquidity provision strategies, have appeared. There are, however, a limited contributions that study the statistical properties of the price, orders, and volumes time series in these markets, such as \cite{Silva_2024}, \cite{wkatorek2024approaching}, and \cite{zhang2024stylized}. Existing research primarily focuses on high-level metrics at coarse granularity, such as daily or hourly prices and returns, offering restricted insights into the fine structure of DEX market dynamics. These earlier works, therefore, miss many of the innovations in the DEX behavior, which are driven by the peculiar trading mechanism and thus only observable at very high-frequency. 

Further, as existing studies primarily focus on price dynamics (which corresponds to the swap events) and none work at very high frequencies, there is, therefore, a significant gap in the literature. Our study fills this gap by focusing on event time analysis, enabling a detailed examination of DEX microstructure dynamics that is characterized by several departures from TradFi. Our work provides novel insights that bridge the gap between the theoretical understanding of traditional markets and the emerging dynamics of DEXs. Specifically, we take the publicly available (on the blockchain) data about the Uniswap v3 market and study its statistical properties at the event level. Our analysis involves both the LTs and LPs activity. We compare the stylized facts found on DEXs with what is observed in TradFi. Next, we show that the main differences are driven by the trading flow organization and the presence of special agents, named Maximal Extractable Value (MEV) searchers, which are involved with the creation of the block of transactions added to the blockchain. Even if this role is formally public, we find that most of the MEV searchers' activity is led by a few agents, thus highlighting the presence of significant technical barriers that prevent small investors from assuming this role. Finally, our work is also aimed at providing help for building realistic simulators of the DEX dynamics at the order level, similarly to what has been done in the last twenty years for CEX.

Overall, our analysis highlights the presence of significant patterns in the data that, while partly paralleling traditional financial markets, also diverge due to the specific market structure of DEXs and their technological infrastructure. These deviations are mostly (though not exclusively) associated with MEV strategies. The main differences are:
\begin{itemize}[label=$\triangleright$]
    \item The individual swap returns are leptokurtic, while, at least for large tick stocks, they are not in TradFi. In swap-time, the kurtosis linearly or super-linearly decays for the first few aggregated swaps, before behaving as a sub-linear power-law as the aggregation scale increases. See Section \ref{subsec:fat_t}.
    
    \item Echo swaps are found in DEXs. Their impact is small yet not negligible, and may be related to MEV attacks occurring on other pools. This is not observed in TradFi. See Section \ref{subsec:echo}.
    
    \item The AutoCorrelation Function (ACF) of returns shows some peaks for small lags that can be associated with echo swaps and the activity of reverse arbitrageurs (at lag 1) and of sandwich attacks (at lags 2 and 3). On the contrary, in TradFi markets, there is only one peak at lag one, which is associated with the bid-ask spread. The arbitrageurs effect is persistent when measuring the ACF in clock-time. We also see a similar pattern in the signed volume ACF. See Section \ref{subsec:ret_acf}.
    
    \item Order flow is correlated, both in swap-time and tick-time, but herding is more important than splitting, contrary to what is observed in TradFi markets. The herding component remains the most relevant even when aggregating transactions from different pools. See Section \ref{subsec:trade_sign}.
    
    \item Liquidity is characterized by significant drops as a consequence of the Concentrated Liquidity mechanism. These drops generate jumps in the marginal price leading to deviations  from the CEX benchmark. Although these divergences are not profitable and may persist for minutes, they  generate artificial volatility in the pools. See Section \ref{sec:liquidity}.
    
    \item There is a very weak correlation between volatility and Liquidity provision range, while in TradFi markets, volatility and spread are strongly correlated. See Section \ref{subsec:range_vs_vol}.
    
    \item The transition probabilities between events display several peaks, which can be associated with MEV strategies. See Section \ref{sec:trans_prob}.
\end{itemize}

The remainder of this work is organized as follows. Section \ref{sec:data} introduces the data we use for this study. Section \ref{sec:mev} analyzes the MEV, which is a major driver of several statistical properties of Uniswap v3. Section \ref{sec:swaps} focuses on the swaps and returns time series. Here, we draw parallels with traditional assets. Section \ref{sec:liquidity} discusses liquidity, corresponding to mint and burn events. Section \ref{sec:trans_prob} examines the transition probabilities between different types of events. Section \ref{sec:conclusion} concludes. Some technical Appendices present descriptive elements that complete the narrative in the main text. Specifically, Appendix \ref{app:uv3md} provides the mathematical formalism of the Uniswap v3 framework. It is intended for readers who are approaching DEXs for the first time. Appendix \ref{app:oep} discusses general patterns about the market activity. Finally, Appendix \ref{app:seqr} shows the numerical results concerning the swap events analysis.
\section{Data}
\label{sec:data}
In this paper, we focus on analyzing the statistical properties and stylized facts of Uniswap v3. As discussed, this choice is driven by the significant volume exchanged on Uniswap v3 and the plethora of data available for carrying out statistical analysis. The study period covers the entire 2023 and 2024. Specifically, we focus on the 24 most active pools\footnote{By ``most active'' we mean those with the largest total number of events (swaps, mints, and burns). Specifically, we consider pools with at least 40,000 events recorded in the study period.}, and involves 15 token pairs. The data are retrieved using the Python package \texttt{eth\_defi} and the official Ethereum subgraph of Uniswap v3 hosted by TheGraph\footnote{Further details on data extraction may be found at \url{https://https://github.com/tradingstrategy-ai/web3-ethereum-defi} and \url{https://thegraph.com/explorer/subgraphs/5zvR82QoaXYFyDEKLZ9t6v9adgnptxYpKpSbxtgVENFV?view=Query&chain=arbitrum-one}, respectively.}. Both methods provide standardized and reliable access to on-chain DeFi information.

To capture a broad spectrum of market dynamics, we categorized the pairs into four distinct groups based on the nature of the tokens involved. For every pair, we specify the numeraire against which the price and returns are computed. The pools are clustered in the same way.
\begin{itemize}[label=$\triangleright$]
\item \textbf{Normal Pairs}: Consisting of a volatile asset and a stablecoin, which serves as the numeraire. The pairs included in this group are \texttt{USDC-WETH}, \texttt{WETH-USDT}, \texttt{WBTC-USDC}, and \texttt{WBTC-USDT}.

\item \textbf{Volatile Pairs}: Comprising of pairs of two volatile assets. The numeraire is the less valuable token, i.e., the one with larger reserves in the pool. The volatile category consists of \texttt{WBTC-WETH}, \texttt{LINK-WETH}, \texttt{MNT-WETH}, and \texttt{UNI-WETH}.

\item \textbf{Stable Pairs}: These pairs consist exclusively of stablecoins and include \texttt{USDC-USDT}, \texttt{DAI-USDC}, \texttt{DAI-USDT}, and \texttt{USDe-USDT}. As all these stablecoins are anchored to the USD, the price of these pairs oscillates around $1$ and is characterized by low volatility. The numeraire used is the second token of each pair.

\item \textbf{Synthetic Pairs}: Defined by pairs that feature WETH or WBTC coupled with ERC-20 wrapped version of liquid staking tokens. The synthetic pairs included are \texttt{WETH-weETH}, \texttt{wstETH-WETH}, and \texttt{WBTC-LBTC}. By construction, the price oscillates around $1$. For each pair, the numeraire is either WETH or WBTC.
\end{itemize}

Henceforth, while the study includes all the pairs above, for illustrative examples we use the \texttt{USDC-WETH} 0.05\% pool, as it is the most traded. When differences between pool clusters are worth highlighting, we display the most traded pool in each cluster: \texttt{USDC-WETH} 0.05\% (Normal cluster); \texttt{WBTC-WETH} 0.05\% (Volatile cluster); \texttt{USDC-USDT} 0.01\% (Stable cluster); \texttt{wstETH-WETH} 0.01\% (Synthetic cluster). Whenever necessary, general quantitative results for all the pools are provided in the appendices.

Figure \ref{img:stf_N_hist} summarizes the amount of data available in the considered pools. There is a significant heterogeneity across pools, especially pertaining to the number of liquidity provision (mint and burn) events. Most of the swaps occur in the Normal pools; the number of LP events is higher in the first two clusters (whose price is not anchored to any fair value) and lower in the last two (whose price oscillates around 1, by design). A richer summary of the data is provided in Appendix \ref{app:oep}, Table \ref{tab:data_summary}. Finally, we refer to Appendix \ref{app:intraday} for an analysis of the pool activity over the course of the day.
\begin{figure}[h]
\centering
    \includegraphics[scale=0.40]{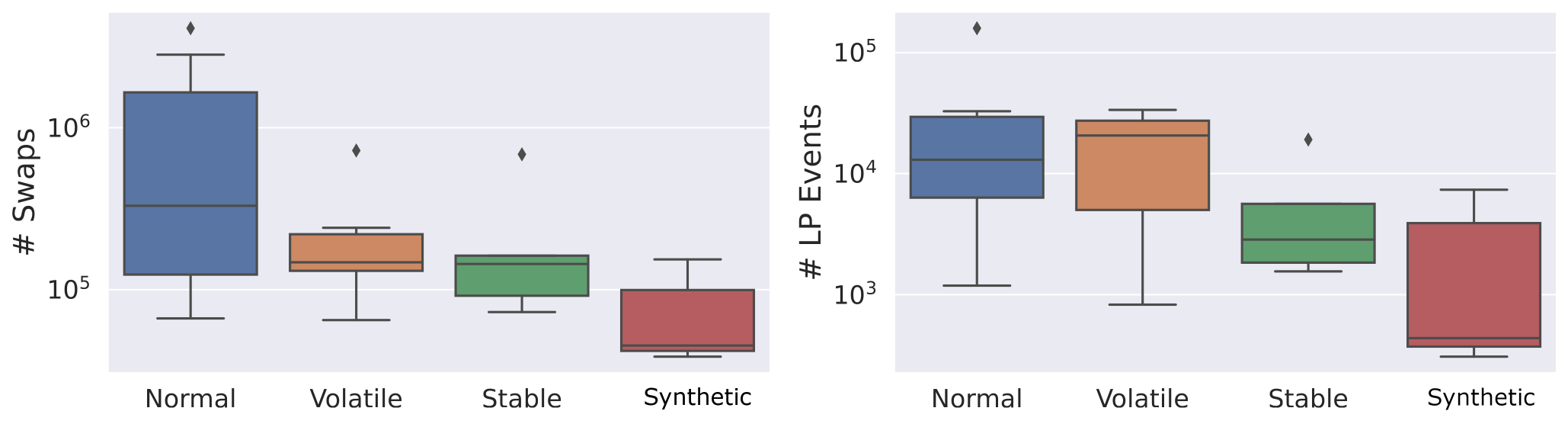}
    \vspace{-.3cm}
	\caption{Amount of available data in the considered period (the entire 2023 and 2024), on log-scale, namely swaps (left) and LP Events (right), which include mints and burns. Each box plot describes all the pools in a cluster.}
\label{img:stf_N_hist}
\end{figure}

We carry out the analysis on different timescales, according to the microstructure property we study. Specifically, we consider:
\begin{itemize}
    \item \textbf{Event-time} increases by one when a new event is recorded in the blockchain. Hereafter, `event' means either a swap, a mint, or a burn occurring in the considered pool. When a new block with multiple events is added to the blockchain, the event time increases by one at each event, and the block index is used to order them.
    
    \item \textbf{Swap, mint, burn-time} refer to filtered versions of event-time, where only one specific type of event is considered. For example, the swap-time counter increases exclusively when a swap event occurs in the pool.
    
    \item \textbf{Tick-time} increases by one when the tick changes. Uniswap v3 price space range is partitioned by discrete ticks. A tick is an integer, defined as $\lfloor \log_{1.0001} S \rfloor$, where $S$ is the marginal price.
    
    \item \textbf{Clock-time} is the usual physical time and we consider different time scales ranging from 30 seconds to 15 minutes.
\end{itemize}
Regarding the tick-time, a brief clarification is necessary. It is well-known that swap events always move the price; however, due to the discrete nature of ticks, they may or may not change in swap-time. Working in tick-time is particularly important when dealing with Concentrated Liquidity. Indeed, a tick movement could change the virtual reserves used to provide liquidity to the swaps, changing the liquidity profile, thus the price formation, and how the fees are distributed among the LPs. Moreover, moving the price tick could also result in higher gas fees (see the white paper \cite{adams2021uniswap} for more details). Finally, interestingly, the magnitude of the scale from swap-time to tick-time is strongly affected by the pair type: For Normal and Volatile pools, the average number of swaps for a tick change ranges between $1$ and $6$. For Synthetic pools, the range is between $11$ and $49$. For Stable pools, the highest recorded average is $131.53$, corresponding to the \texttt{DAI-USDC} 0.01\% pool. We suggest that this outlier is due to the DAI value being anchored to the USDC via both collateralization and a pegging mechanism.
\section{Maximal Extractable Value}
\label{sec:mev}

Before discussing the statistical analysis, we provide some background on the technical organization of DEXs and, specifically, on how transactions are submitted and validated on the Ethereum blockchain. This peculiar framework allows agents charged with handling orders to extract profit from transaction manipulations such as rearrangements, additions, or removals. This profit is known as Maximal Extractable Value (MEV)\footnote{Historically, it has been introduced as Miner Extractable Value, see \cite{daian2019flash}.}, and it is 
an important driver of the trading dynamics in DEXs and thus in Uniswap. In this paper, we show that the pervasiveness of MEV practices is a key driver of the stylized facts of price and orders in DEX.

In TradFi, orders are instantaneously processed, and the difference between submission from the user and insertion in the limit order book is usually of the order of microsecond, and it is rare to find trades sharing the same timestamp. Contrastingly, the order flow in DEXs is lumpy, as the transactions are grouped into blocks that are validated sparsely in time -- currently, on the Ethereum blockchain, every $12$ seconds. Hence, submitted orders are not immediately executed, but rather, they need to wait before being added to the blockchain.

\begin{figure}[h]
    \centering
    \includegraphics[width=1\linewidth]{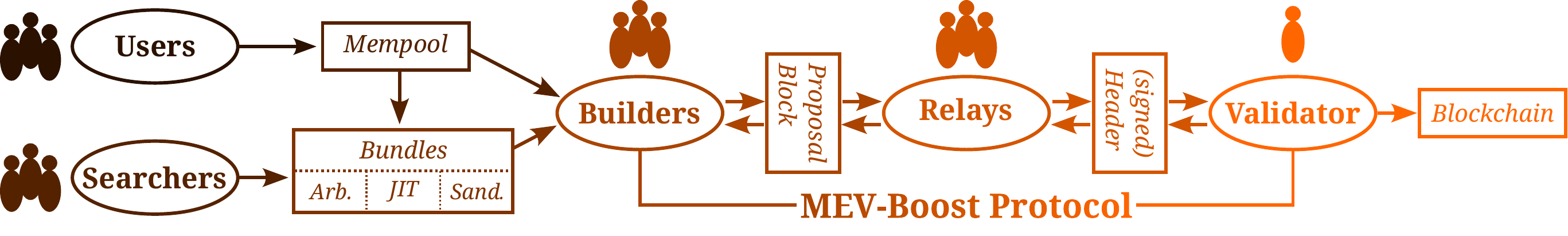}
    \caption{DEX orders submission pipeline. Users submit their orders to the mepool, which is accessed by searchers that create bundles of transactions to extract MEV. Each builder creates its own block proposal by taking transactions from the mempools and bundles from the searchers. Then, builders submit their proposals to the validator, which selects the most profitable one to be added to the blockchain. Relays act as intermediaries, and the whole auction is handled by the MEV-Boost protocol.}
    \label{fig:dex_pipeline}
\end{figure}

This is the main source of differences in the stylized facts between DeFi and TradFi, especially at event time. The orders submitted between two blocks are received by an Ethereum node and stored in the `memory pool' (mempool), which can be seen as a `waiting room' where transactions stay before being validated and appended to the next block.
Typically, the mempool is public and accessible to everyone\footnote{For example, via Ethernow \url{https://ethernow.xyz/mempool/all}.}. Before private bundle auctions \cite{flashbotsAuctionOverview} and later MEV-Boost implementation of Proposer-Builder Separation (PBS) \cite{mevBoost} by Flashbots, \textit{MEV searchers} often competed through public Priority Gas Auctions (PGAs) \cite{FlashBoys}. As of 2021, Flashbots’ MEV-Geth moved MEV bidding to sealed-bid private bundles that a large share of miners adopted, reducing PGA spam in the public mempool that contributed to gas fee price spikes. MEV-Boost then standardized PBS on proof-of-stake Ethereum by letting validators source blocks from specialized actors, the \textit{block builders} \footnote{While PGAs contributed to fee spikes, high gas periods also reflected general demand; moreover, the Ethereum Improvement Proposal (EIP) 1559 reformed the fee market and dampened fee-volatility dynamics \cite{eip1559}},
thus, introducing a new marketplace where MEV opportunities could be captured off-chain. The purpose was, instead of competing in the public mempool, to allow MEV searchers, trying to extract MEV by continuously scanning the mempool, to bundle transactions and send them directly to the block builders in a bid-auction mechanism. Then, builders construct block proposals that are submitted to a validator, again in a bid-auction mechanism, which is managed by a trusted third-party known as \textit{relay}. Figure \ref{fig:dex_pipeline} summarizes the pipeline. For a detailed description of this chain of actors, see \cite{oz2024wins}.
\\

For our purposes, MEV searchers play a central role. Indeed, they have significant freedom pertaining to which transactions to include in the bundles and what ordering they have -- which is generally different from the arrival order in the mempool. Thus, if they find exploitable situations (e.g., a big swap to front-run), they can rearrange the transactions to gain a profit.
The most widely used MEV strategies include\footnote{For a comprehensive survey about MEV in DeFi see \cite{materwala2024maximal}.}:
\begin{itemize}
    \item \textbf{Arbitrage opportunities}: MEV searchers can exploit price mismatches between two DEXs or between a DEX and a CEX. Consequently, they contribute to market efficiency.
    
    \item \textbf{Liquidations}: DeFi protocols allow users to borrow money against collateral. If the collateral's value drops below a threshold, the user faces liquidation, where their crypto assets are sold to repay the debt. Liquidators buy these assets at a discount, making a risk-free profit. Consequently, MEV searchers compete for these opportunities.
    
    \item \textbf{Sandwich attacks}:
    strategies designed for gaining from the price impact of a large swap (victim) pending in the mempool. The MEV searcher first places a swap in the same direction as the one of the victim is placed (also known as front run); then, the victim swap is executed; lastly, the attacker closes the position by placing a swap with an equal amount and opposite sign to the front run (also known as back run).
    
    \item \textbf{Just-in-Time (JIT)} liquidity \cite{wan2022just}:
    a strategy for gaining the fee that a large swap in the mempool will pay once executed. The MEV searcher deposits liquidity in a very narrow price range (containing the current price) just before the transaction is processed, earning most of the trading fee, and removes the liquidity immediately after the transaction. This strategy is also referred to as \textit{LP sandwich} or \textit{active liquidity provision}\footnote{Despite this definition is very appealing, it should not be confused with the so-called \textit{active liquidity}, that is the sum of all the liquidity whose range contains the price, as discussed in Appendix \ref{app:subsec_uv3}.} \cite{capponi2023paradox}, as the liquidity provision is elicited by a pending order in the mempool. 

    \item \textbf{Mixed (Sandwich+JIT) attacks}: When a large swap is detected in the mempool, a mixing attack made up of a JIT encapsulated into a sandwich could sometimes bring larger profit. Specifically, the exact scheme is: front run, mint liquidity, victim swap, burn liquidity, back run.
\end{itemize}
To summarize, there are mainly three strategies that an MEV searcher uses to react to a large pending swap. The choice strongly depends on the liquidity in the pool and the chance for hedging when providing active liquidity. Nonetheless, the key parameter is the pool fee. In low-fee pools, performing sandwich attacks is cheaper, and providing active liquidity provision is less profitable; in high-fee pools, the opposite is true. In the following subsections, we discuss some MEV strategies in more detail and study their prevalence in the investigated dataset.

\subsection{Just-in-Time Liquidity}
\label{subsec:jit_liq}
Liquidity provision is classically thought of as a passive operation, as the LP puts tokens in the pool and waits for LTs to enter the market. In contrast, Just-in-Time liquidity \cite{wan2022just} is a kind of active liquidity provision \cite{capponi2023paradox} aimed at zeroing the periods when the LP's tokens are unused and reducing the impact of adverse selection. When a big swap appears in the mempool, a MEV searcher can choose to instantaneously provide liquidity to that order. The attacker's tokens are used and freed in the same block, and he/she earn a large fraction of the fee paid by the LT\footnote{In our dataset, the median share of the swap fee captured by the JIT attacker (computed over all the JIT attacks within a pool) exceeds 90\% for most of the pools.}. Specifically, the structure that we observe in the block is made up of a mint order with a very narrow tick range\footnote{Usually, the range is the narrowest possible. That is, the difference between the upper and lower tick is equal to the so-called \texttt{tickSpacing}, see \cite{adams2021uniswap} for further details.} followed by the victim swap (or more than one) and, finally, the burn order that removes the previously minted liquidity, i.e., the mint and burn are placed by the same wallet and have the same liquidity amount and range. Algorithm \ref{alg:jit_detection} shows the pseudo-code for detecting JITs occurring in a pool.

\begin{algorithm}[H]
\caption{Just-In-Time liquidity detection}
\label{alg:jit_detection}
\begin{algorithmic}[1]
\State \textbf{Input:} \texttt{pool} Dataframe.
\State \textbf{Output:} \texttt{JITs} list of triples describing JIT attacks.

\State Initialize \texttt{JITs = list()}
\For{\texttt{block} in \texttt{pool.groupby("block\_number")}}
    \For{\texttt{mint} in \texttt{block[ event=="Mint" ]}}
        \If{exists \texttt{burn} in \texttt{block[ event=="Burn" ]} s.t. (\texttt{burn.log\_index} > \texttt{mint.log\_index}) and (\texttt{burn.values}== \texttt{mint.values})}
            \State Define \texttt{swaps = block[ event=="SwapX2Y" or event=="SwapY2X" ]}
            \If{\texttt{swaps not empty}}
                \State \texttt{JITs.append( (mint, burn, swaps) )}
            \EndIf
        \EndIf
    \EndFor
\EndFor
\end{algorithmic}
\end{algorithm}

Specifically, our JITs detection algorithm works as follows:
\begin{enumerate}
    \item The input is a dataframe, \texttt{pool}, whose rows are the events occurring in the pool. The columns of the dataframe are: \texttt{block\_number} and \texttt{log\_index} indicate the block and the position of the event; \texttt{wallet} is for the wallet address sending the transaction; \texttt{event} is for the event-type -- either \texttt{SwapX2Y}, \texttt{SwapY2X}, \texttt{Mint}, or \texttt{Burn}; \texttt{tick\_l}, \texttt{tick\_u}, and \texttt{amount} are for the lower tick, upper tick, and virtual liquidity of the liquidity position -- they are \texttt{nan} if \texttt{event} is \texttt{SwapX2Y} or \texttt{SwapY2X}. See \textbf{line 1}.
    \item The output is a list, \texttt{JITs}, containing the JIT attacks. These are recorded as triples: the first element is the attacker mint, the second element is the attacker burn, and the third element is the victim swaps. See \textbf{lines 2-3}.
    \item Iterate over the blocks. For every block, scan all the mint and burn events. Looks for the couples \texttt{(mint, burn)} such that:
        \begin{itemize}
            \item \texttt{mint} is executed in the same block and before \texttt{burn};
            \item \texttt{mint[tick\_l, tick\_u, amount, wallet]} is equal to \texttt{burn[tick\_l, tick\_u, amount, wallet]}.
        \end{itemize}
    See \textbf{lines 4-6}.
    \item If \texttt{mint} and \texttt{burn} are spaced by some \texttt{swaps}, then save the triple \texttt{(mint, burn, swaps)} as a JIT attack. See \textbf{lines 7-9}.
\end{enumerate}

It is worth noting, unlike in \cite{wan2022just}, we do not impose the constraint that the three events are consecutive in the block and instead only require the log index order to be mint, swaps, and burn. That is, we also identify as JITs the patterns that are spaced out by other transactions within the same block. This is because the occurrence of other events, in different pools and/or blockchain protocols, inserted between does not harm the economic reason and the profitability of JIT attacks.

JIT attacks usually have a beneficial effect on the LT\footnote{There is a slight misrepresentation in calling the user that places the central swap in a JIT attack ``victim''. Nonetheless, we use the definition ``victim'' to unify JIT and sandwich nomenclature and to be consistent with the notation used in the MEV literature.}, as the swap price impact is reduced due to the presence of new liquidity in the pool.
Finally, even if the LT is formally trading against the pool, as most of the liquidity is provided by the MEV searcher, the LT is mainly trading against the attacker. Thus, the searcher is in a similar position as the dealer in CEXs, as the searcher is positively (negatively) exposed on the token sold (bought) by the LT. Thus, the searcher faces the classical internalization-externalization problem (e.g., see \cite{barzykin2022market}), where the problem is whether to hedge this exposure on other markets (either DEX or CEX) or hold the exposure and wait for other MEV opportunities to offset it.

\begin{figure}[h]
	\centering
	\includegraphics[scale=0.35]{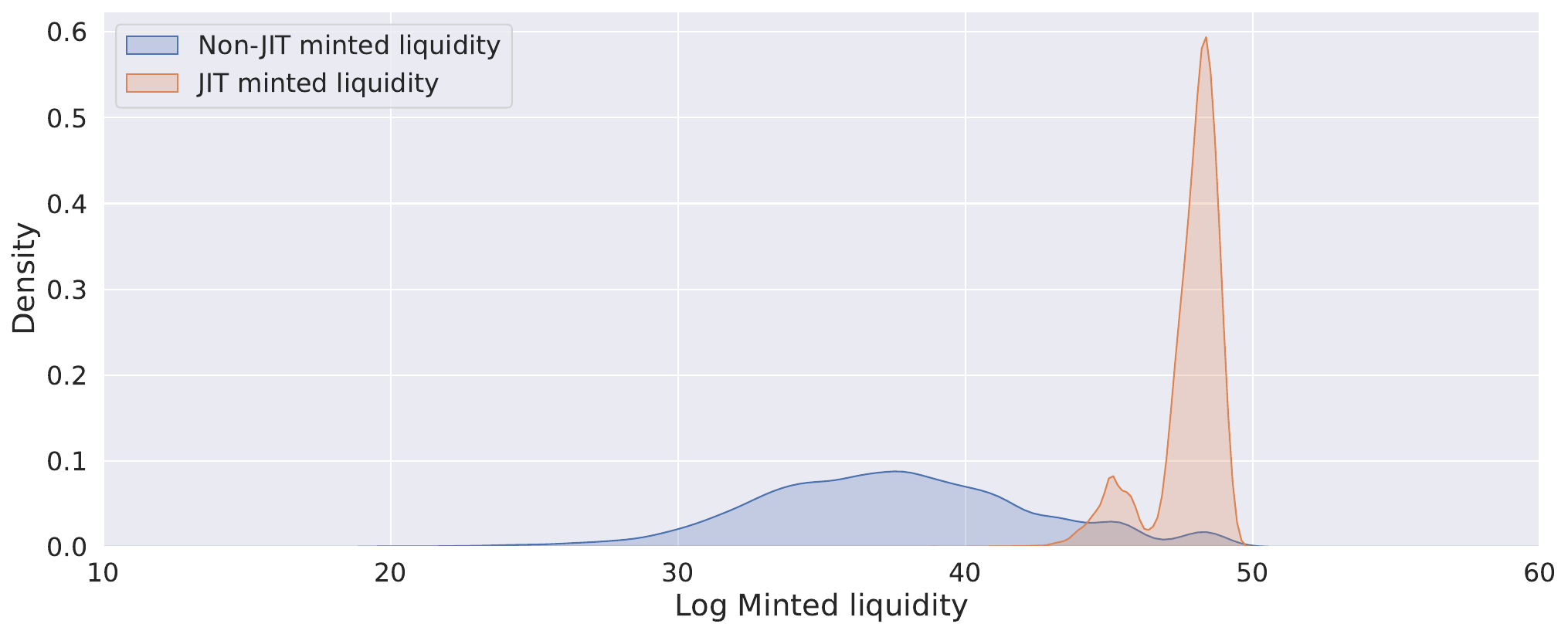}
	\caption{Decomposition of log liquidity minted in JIT liquidity and non-JIT liquidity for the \texttt{USDC-WETH} 0.05\% pool. JIT liquidity is concentrated in the top region of the abscissa. This is mainly due to the narrow range characterized by these events.}
	\label{img:jit_liq_vs_nonjit_liq}
\end{figure}

As the JIT attacker aims to take almost all the fee paid by the victim LT, the virtual liquidity\footnote{See Appendix \ref{app:subsec_uv3} for further details on the virtual liquidity in Uniswap v3} minted by the active LP is expected to be far larger than the median liquidity minted by passive LPs. This is indeed the case, as shown in Figure \ref{img:jit_liq_vs_nonjit_liq}, which compares the Kernel Density Estimation (KDE) of the liquidity minted by passive and active LPs for the \texttt{USDC-WETH} 0.05\% pool. The latter is significantly larger as the attacker, being focused on a single swap, chooses a very narrow tick range.

\begin{figure}[t]
    \centering
    \includegraphics[height=0.17\linewidth]{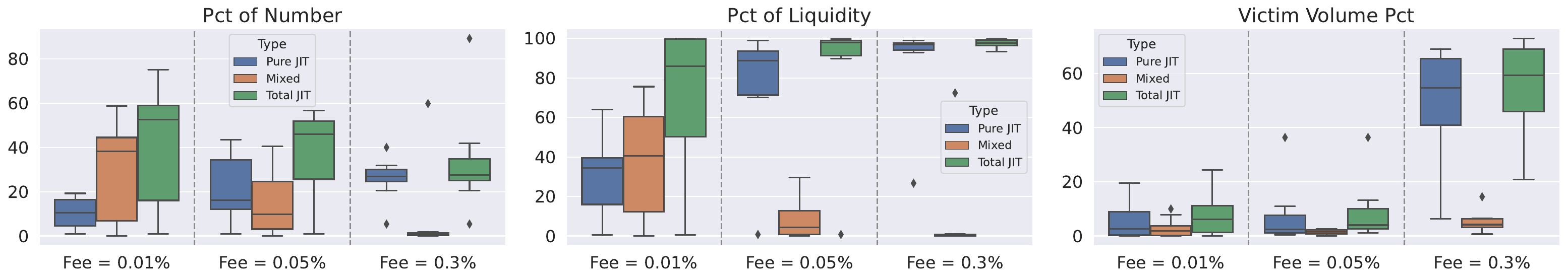}
    \caption{JIT impact -- The left panel shows the percentage of mint events that are JIT; the central panel shows the percentage of minted liquidity belonging to JIT attacks; the right panel shows the percentage of the total volume traded in the pool that has been victim of a sandwich attack. In each panel, the information belonging to pool with different fee tiers is shown in a different box. Moreover, blue boxes show pure JIT events; orange boxes are for JIT involved in mixed JIT-sandwich attacks; green boxes show the total (pure+mixed) JIT attacks.}
    \label{fig:enter-jit_impact_x_paper}
\end{figure}

Figure \ref{fig:enter-jit_impact_x_paper} shows the prevalence of JIT strategies in the pool as the percentage of mint events and minted liquidity involved in JIT attacks, as well as the percentage volume of victim swaps.
We find that more than 20\% of the total mint operations and more than 50\% of the minted liquidity in our data is due to JIT\footnote{The percentage jumps to 90\% when considering fee tiers 0.05\% and 0.3\%. However, it is slightly biased by the fact that the JIT ranges are usually very narrow.}. This seems to be consistent with the active liquidity provision as proposed in \cite{capponi2023paradox}, as these events appear more akin to a fundamental component of the market rather than a rare profit opportunity for MEV searchers. Furthermore, different fee tiers behave significantly differently. Indeed, the empirical observations confirm what has been previously observed: pure JIT attacks are profitable almost only in high-fee pools.

We next investigate the population of active liquidity providers. A previous analysis carried out in \cite{xiong2023demystifying} highlights that more than $70\%$ of all JITs have been made by a single user. However, the dataset used spans from June 2021 to January 2023; thus, only January 2023 overlaps with our dataset. Thus, following along the lines of  \cite{xiong2023demystifying}, we study what occurred in the next two years. We collect the wallet addresses of the active LPs and compute the percentage of transactions executed from each wallet. Interestingly, the situation seems to be less heterogeneous. Nonetheless, $26\%$ of all the JITs were carried out by an MEV bot operated by a user publicly known as \texttt{jaredfromsubway} \footnote{Wallet address \texttt{0xae2Fc483527B8EF99EB5D9B44875F005ba1FaE13}}, one of the most famous MEV searchers on the Ethereum blockchain. Overall, there are $13$ accounts with at least $2\%$ of the transactions. This number rises to $25$ and $42$ if we set the threshold to $1\%$ and $0.5\%$, respectively. Ultimately, the trend seems to be an opening of the JIT provider role to even more agents.

Contrastingly, mixed attacks appear far more centralized. \texttt{jaredfromsubway} is still the most active wallet, responsible for about $88\%$ of them. Only $3$ attackers contribute more than $1\%$ and $2\%$; and, $5$ searchers contribute more than $0.5\%$. 

\subsection{Sandwich Attacks}
\label{subsec:sand_att}
Sandwich attacks are performed by MEV searchers willing to gain from the price impact of a large swap pending in the mempool. To achieve this task, the searcher adds their own swaps just before and after the victim's, with the intention of front-running and then back-running it.
It is worth noting that this strategy has a negative effect on the LT, as the front run moves the price unfavorably, leading the victim to buy (sell) at a higher (lower) price. Algorithm \ref{alg:sandwich_detection} shows the algorithm we use for detecting these attacks.
\begin{algorithm}[H]
\caption{Detection of Sandwich Attacks}
\label{alg:sandwich_detection}
\begin{algorithmic}[1]
\State \textbf{Input:} \texttt{pool} Dataframe.
\State \textbf{Output:} \texttt{Sands} list of triples describing sandwich attacks.

\State Initialize \texttt{Sands = list()}
\For{\texttt{block} in \texttt{pool.groupby("block\_number")}}
    \State Define \texttt{swaps = block[ event=="SwapX2Y" or event=="SwapY2X" ]; N = len(swaps)}
    \For{\texttt{front} in \texttt{swaps.pop()}}
        \If{front.event == "SwapY2X"}
            \State{\texttt{back = swaps[ event=="SwapX2Y" ][0]}}
        \Else
            \State{\texttt{back = swaps[ event=="SwapY2X" ][0]}}
        \EndIf
        \If{(\texttt{back not nan}) and (\texttt{back.wallet == front.wallet})}
            \State Define \texttt{victims = swaps[ front.log\_index < log\_index < back.log\_index ]}
            \If{\texttt{victims not empty}}
                \State \texttt{Sands.append( (front, back, victims) )}
            \EndIf
        \EndIf
    \EndFor
\EndFor
\end{algorithmic}
\end{algorithm}

In detail, our algorithm works as follows:
\begin{enumerate}
    \item The input is the same dataframe \texttt{pool} as in Subsection \ref{subsec:jit_liq}. See \textbf{line 1}.
    \item The output \texttt{Sands} is a list of triples, each one describing a sandwich attack: (front-run, back-run, one or more victims). See \textbf{lines 2-3}.
    \item Iterate over the blocks. Scan all the swaps in a block as potential front-run. See \textbf{lines 4-6}.
    \item Search for the first swap in the block with the opposite sign. If it exists and has the same wallet as the candidate front-run\footnote{It is possible that the attacker uses two different wallets. For example, the searcher could move the output of the front-run to another wallet before the back-run. As an additional gas fee should be paid for this, it is not very clear if this strategy makes sense. Nonetheless, the presence of two-wallet attacks could explain the inconsistency between the number of sandwiches found by different sources, which is highlighted in \cite{li2024geth}. Thus, we left this kind of analysis for further research.}, then it is a potential back-run. See \textbf{lines 7-12}.
    \item If the potential attacker's swaps are spaced by at least one victim swap of the same sign as the front-run, then classify the pattern as a sandwich attack. See \textbf{lines 13-15}.
\end{enumerate}

\begin{figure}[h]
	\centering
	\includegraphics[width=1.0\textwidth]{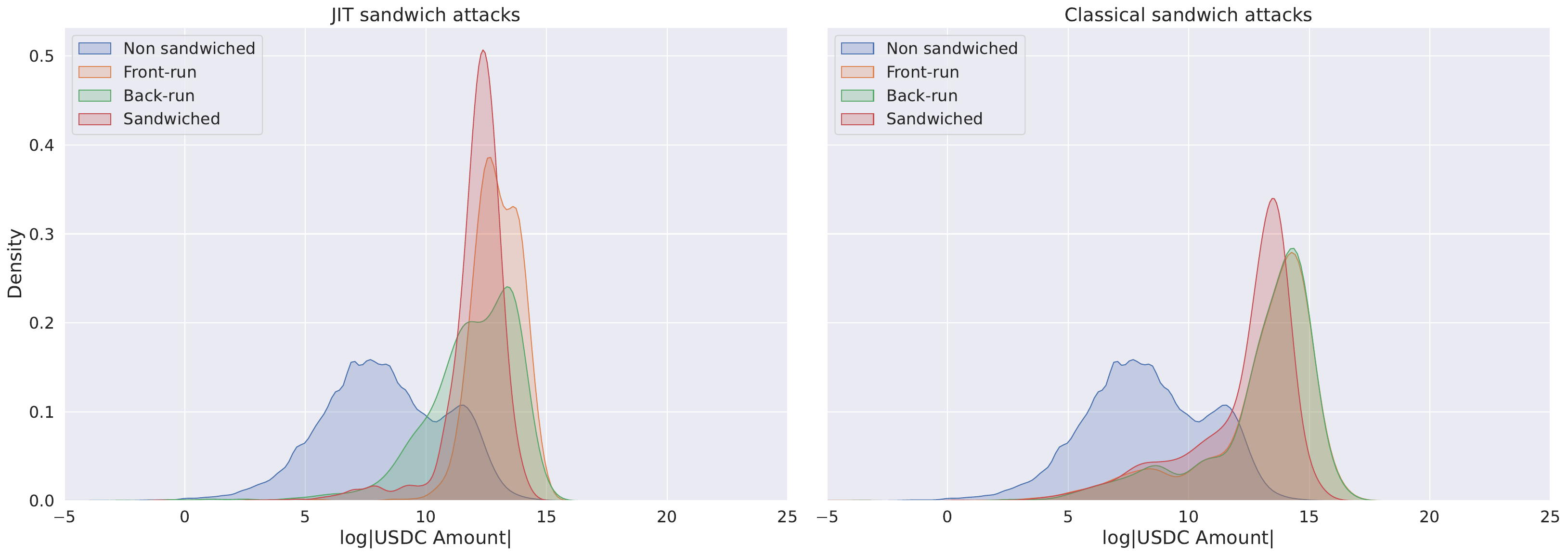}
	\caption{Comparison of KDE of the absolute value of the log USDC amount swapped during all the sandwich attacks discovered ($36,353$) on \texttt{USDC-WETH} $0.05\%$. The left plot is for mixed sandwich+JIT attacks; the right plot is for pure sandwich. The blue area is for swaps not involved in sandwich attacks (non-sandwiched swaps); the red area is for the victim swaps; the orange and green areas are for the front-run and back-run swaps.}
	\label{img:sandwich}
\end{figure}

After applying our procedure, sandwich attacks targeting a single intermediate swap account for $\sim$95\% of all cases, followed by attacks with two swaps ($\sim$4.5\%), with the remaining patterns in total representing less than $1\%$ of the cases. The largest number of swaps affected by a single attack is $11$. Figure \ref{img:sandwich} shows the KDE of the logarithm of the absolute value of the USDC amount swapped in the sandwich attacks identified on the \texttt{USDC-WETH} $0.05\%$ pool.
There is a significant deviation in the amounts involved with sandwich attacks and the other swap amounts. This is expected, as the larger the victim's and attacker's swaps, the higher the attacker's profit. Furthermore, the front-run and back-run amounts in the pure sandwich attacks are almost equal (as it happens in TradFi), as they are designed to eliminate all exposure. Instead, in mixed attacks, the back-run distribution is shifted to the left of the front-run one. This is because part of the exposure opened with the front-run is liquidated trading against the victim in the JIT part of the attack.

\begin{figure}[h]
    \centering
    \includegraphics[height=0.17\linewidth]{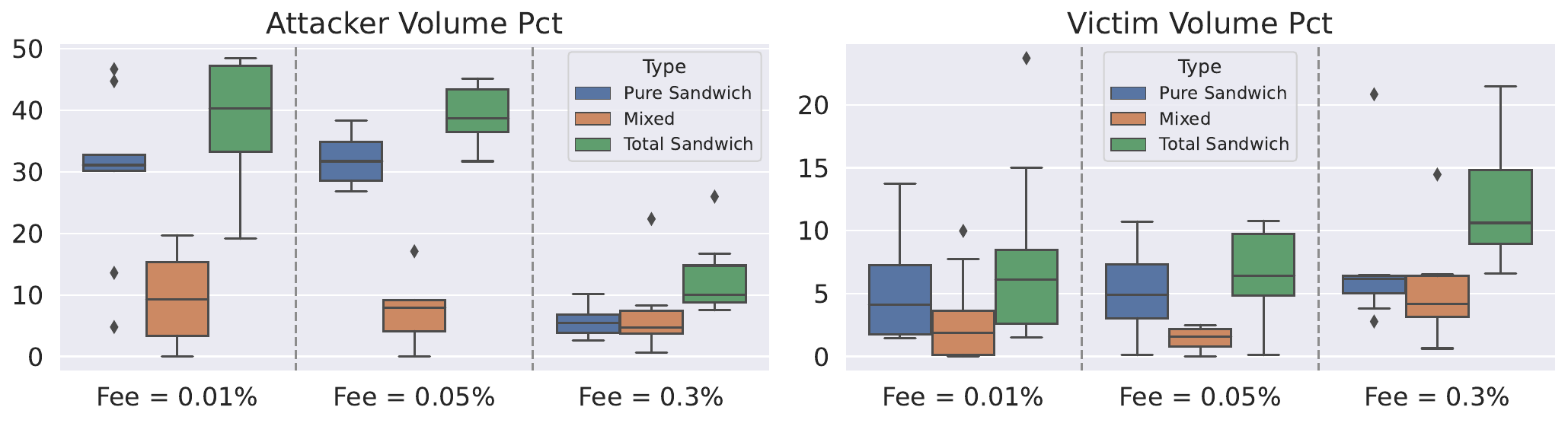}
    \caption{Sandwich impact -- The left plot shows the percentage of the total volume traded that is involved with sandwich attacks (front-run and back-run). The right plot shows the percentage linked to victims. Blue and orange boxes are for pure and mixed attacks, green boxes are for their sum. Pools have been grouped according to the fee tier.}
    \label{fig:sand_impact_x_paper}
\end{figure}

Sandwich attacks account for a sizable fraction of the volume traded in Uniswap v3. Figure \ref{fig:sand_impact_x_paper} shows the percentage volume related to sandwich attacks, both pure and mixed. Surprisingly, in low-fee tier pools, about $40\%$ of the total traded volume is related to malignant sandwich swaps. When the fee tier increases, however, pure sandwich attacks become less frequent, as they become more expensive and JIT strategies become more rewarding. From this perspective, the pool fee can be intended as the price of the protection against sandwich attacks.

Regarding the attackers' identity, we find that  $\sim64\%$ of the pure sandwich attacks have been carried out by \texttt{jaredfromsubway}. Beyond this, $5$, $8$, and $24$ agents contribute at least $2\%$, $1\%$, and $0.5\%$ of the total attacks, respectively.

\subsection{Echo Swaps}
\label{subsec:echo}
\begin{figure}[h]
    \centering
    \includegraphics[scale=0.45]{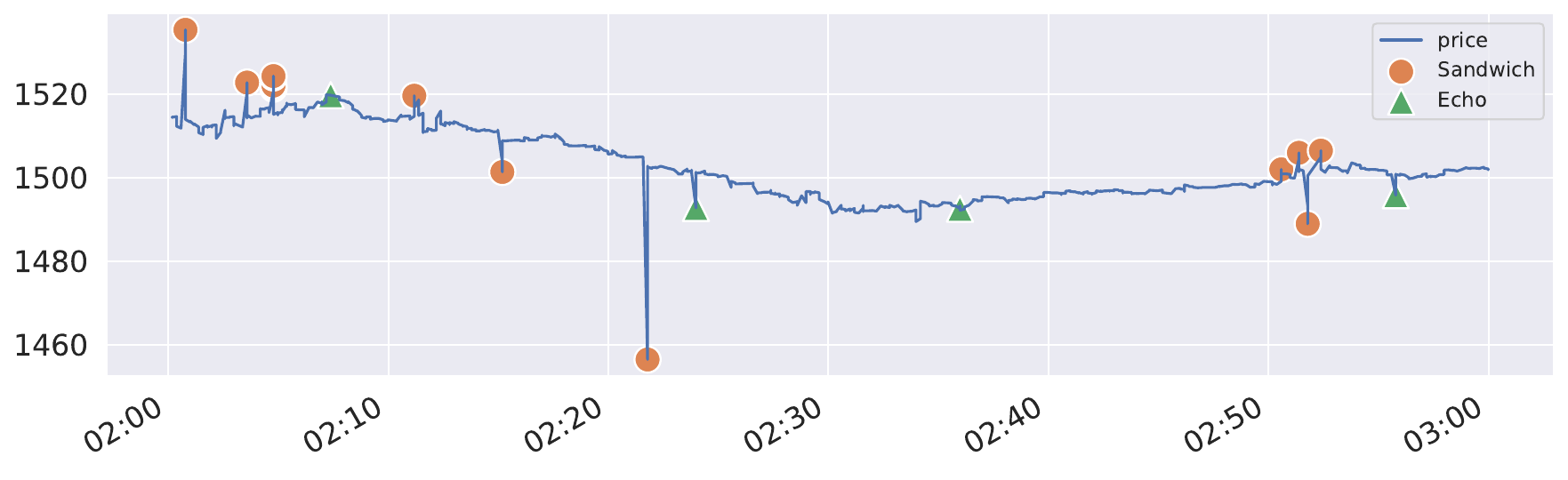}
    \caption{\texttt{USDC-WETH} 0.05\% marginal price (swap-time) on March 11, 2023, between 2 and 3 am (UTC). Orange circles mark sandwich attacks; green triangles are for echo swaps.}
    \label{fig:example_peaks_echo}
\end{figure}

To conclude the discussion about MEV strategies, it is worth mentioning a peculiar pattern that we name \textit{echo swaps}. It consists of two swaps with opposite signs and placed by the same agent in the same block, and without any other swaps in between. In other words, they can be thought of as sandwich attacks without any victim swaps in between. Echo swaps and sandwich attacks typically appear as sharp peaks in the price time series observed in swap-time. Figure \ref{fig:example_peaks_echo} shows an example from the \texttt{USDC-WETH} $0.05\%$ pool.

\begin{figure}[h]
    \centering
    \includegraphics[height=0.17\linewidth]{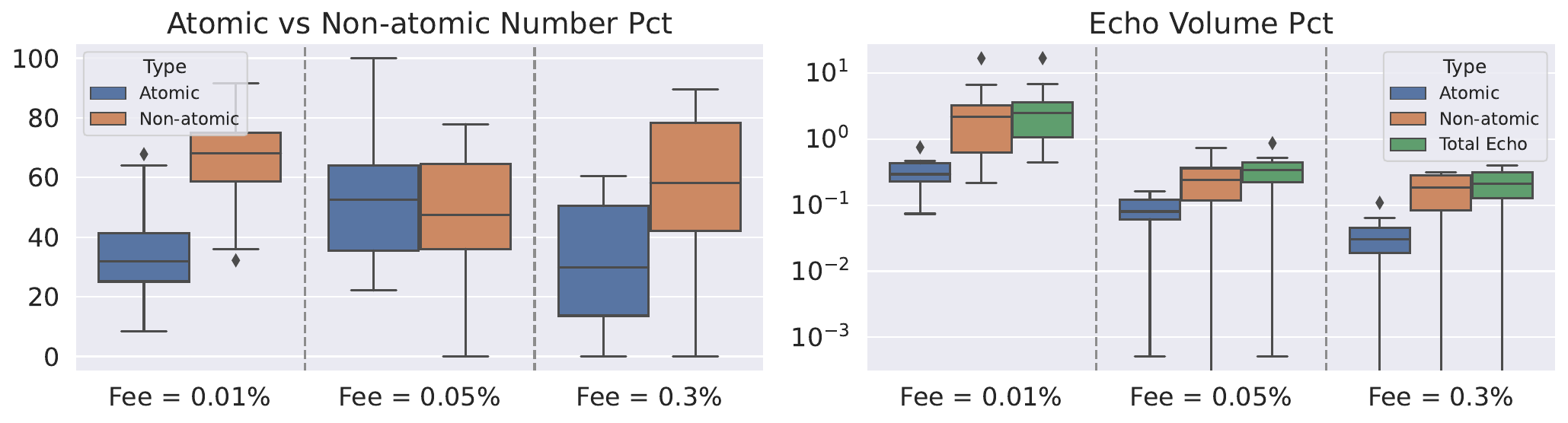}
    \caption{Echo Swaps impact -- The left plot shows the percentage of echo swaps that are atomic (i.e., the swaps have the same transaction hash) or non-atomic. Blue (orange) boxes are for (non) atomic echo swaps. The right plot displays the percentage of the total volume swapped in the pools that is associated with echo swaps. Green boxes are for the sum of atomic and non-atomic echo swaps.}
    \label{fig:echo_impact_x_paper}
\end{figure}

Regarding the transaction hashes of the swaps, sometimes they are equal, meaning echo swaps are part of an atomic transaction, and other times they are different. Figure \ref{fig:echo_impact_x_paper} shows the statistics related to echo swaps. We find no specific pattern in the percentage of atomic and non-atomic echo swaps. Regarding volume, their impact is usually smaller than $1\%$ of the total swapped volume ($17$ out of $24$ pools), and only in the \texttt{USDC-USDT} $0.01\%$ and \texttt{DAI-USDC} $0.01\%$ pools is it larger than $5\%$.

The economic motivations underlying echo swaps remain uncertain, and we leave a detailed exploration of these drivers to future research. However, our current conjecture is that they are used to supply the LT with tokens that are needed for MEV positions in other pools.

By analyzing their transaction hashes, we find that about $65\%$ of them involve more than one pool.

Regarding the identity of the wallets behind echo swaps, we found that the most active one is still \texttt{jaredfromsubway} (8\%). Generally, echo swaps are quite spread, with $8$, $15$, and $34$ agents responsible for at least $2\%$, $1\%$, and $0.5\%$ of their total, respectively. 

\begin{figure}
    \centering
    \includegraphics[width=0.5\linewidth]{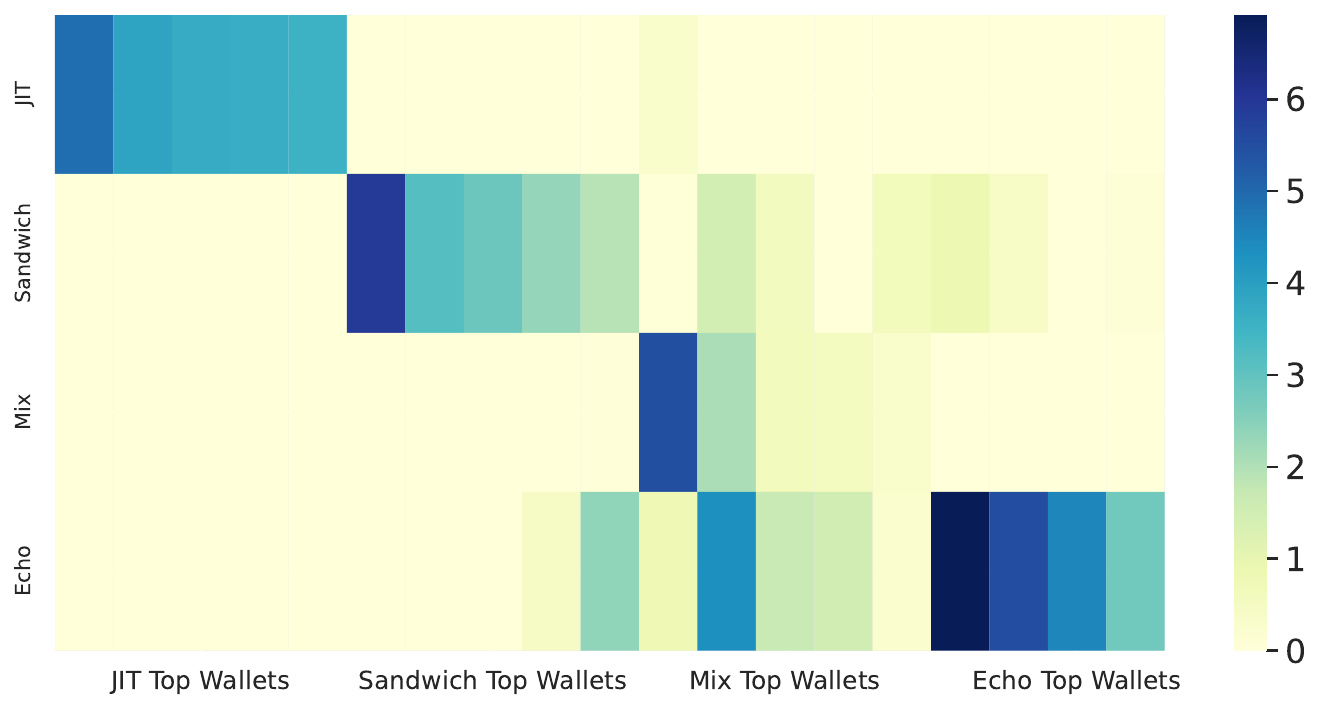}
    \caption{Percentage MEV activity from the most active attackers of each strategy (\texttt{jaredfromsubway} \texttt{0xae2Fc483527B8EF99EB5D9B44875F005ba1FaE13} excluded from the computation). On the x-axis, there are the top 5 attackers of each strategy. On the y-axis, there are the percentages of MEV attacks carried out.}
    \label{fig:mev_attackers_spec_x_paper}
\end{figure}

Finally, we investigate whether attackers `specialize', i.e., do agents perform one type of MEV strategy and neglect the other MEV opportunities. Our findings are shown in Figure \ref{fig:mev_attackers_spec_x_paper}. The most striking is that JIT attackers appear to be separated from the others. Further, we note a dependence between searchers of different strategies. Specifically, sandwich and mix wallets are usually also responsible for echo patterns. Similarly, mix and echo wallets also perform sandwich attacks. We interpret these results as evidence that echo swaps are related to MEV strategies, as they are often executed by wallets known to be MEV searchers.

\section{Statistics of the Swaps}
\label{sec:swaps}

In this section, we study some stylized facts about the swap events and the return time series. Clearly, these quantities are strictly related, as swaps are the only events that can directly modify the price. As shown in Figure \ref{img:stf_N_hist}, swaps represent the majority of the events, accounting for 90\% to 99\% of the entire dataset, depending on the pool. Moreover, as MEV strategies are responsible for a huge share of the swapped volume (see Figure \ref{fig:sand_impact_x_paper}), they drive some peculiar patterns in the return time series observed in swap-time. In the following, we consider log-returns. Table \ref{tab:SrL_summary} in Appendix \ref{app:oep} summarizes the main information about the returns over different pools.
\subsection{Fat-Tail Return Distribution}
\label{subsec:fat_t}
Fat tails are ubiquitous across returns in financial markets. It is well known that the main driver of prices in DEXs is activity in CEXs \cite{cartea2023decentralised}. Nonetheless, over very short time scales (e.g., one or a few swaps), the processes are not synchronized, making the resulting return distribution uncertain. Therefore, we investigate the Gaussianity of the return distribution through the D'Agostino-Pearson's test \cite{d1973tests}. The test, applied both in swap-time and tick-time, leads to the rejection of the null hypothesis of Gaussian returns. Contrastingly, if we aggregate more swaps, then the DEX price dynamics should mimic the CEX dynamics. It is known that by aggregating more and more returns they become more and more Gaussian.

\begin{figure}[H]
	\centering
	\includegraphics[scale=0.10]{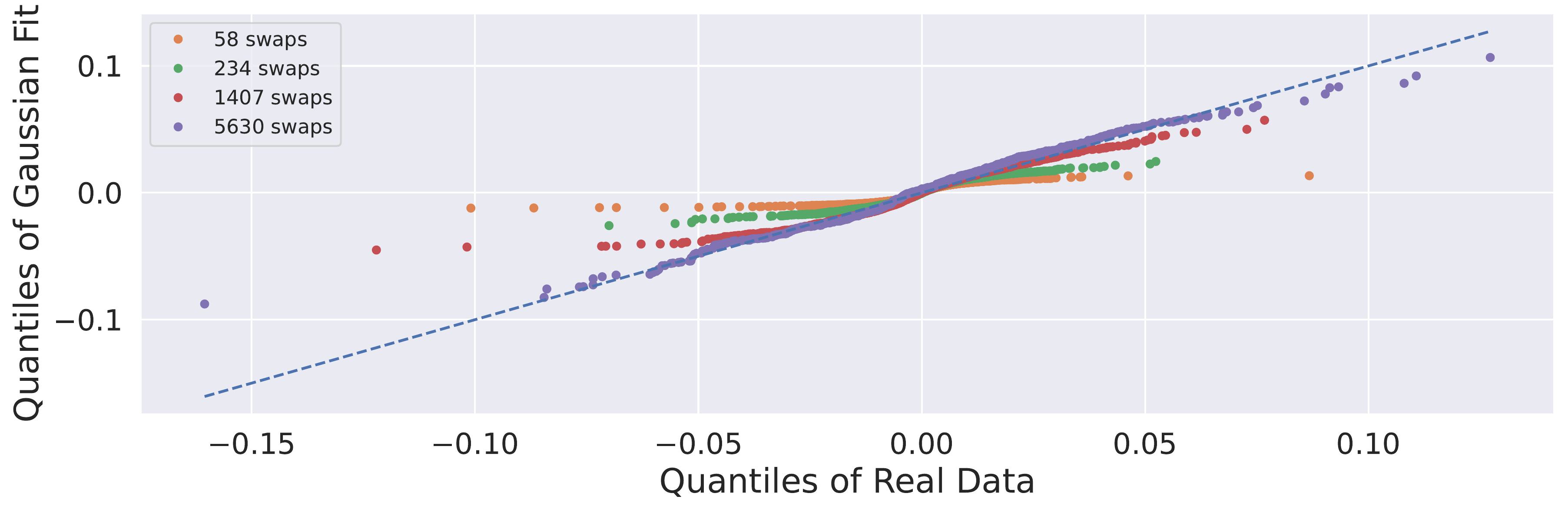}
	\caption{Q-Q Plot of the \texttt{USDC-WETH} 0.05\% returns - empirical distribution vs Gaussian fit at different levels of swaps aggregation. 58, 234, 1,407, and 5,630 are the average swaps in 15 minutes, 1 hour, 6 hours, and 1 day, respectively.}
	\label{img:stf_fat_tails_event_agg_usdc_weth_005}
\end{figure}

Convergence to Gaussian occurs more quickly when working with transactions compared with working with clock time \cite{chakraborti2011econophysics}. This observation may be partially replicated in Uniswap by generating a Q-Q plot of empirical data and the Gaussian fit as the number of aggregated blocks increases. Specifically, we aggregate $n$ swaps with $n$ equal to the average number of swaps in 15 minutes, 1 hour, 6 hours, and 1 day. The result of this exercise, for the \texttt{USDC-WETH} 0.05\% pool, is reported in Figure \ref{img:stf_fat_tails_event_agg_usdc_weth_005}. The figure indicates that as the number of aggregated swaps increases, the Q–Q plot progressively aligns with the diagonal, suggesting convergence toward a Gaussian distribution.

The non-Gaussianity of returns can be examined through their kurtosis. At long aggregation scales, we expect DEX and CEX prices to be synchronized. When aggregating only a few swaps or blocks, however, the DEX dynamics can diverge from those of CEX. Focusing on the returns' excess kurtosis in swap-time, we find extremely large values (typically decreasing as the fee tier $\phi$ increases), almost always above $1,000$. The source for this pattern is not restricted to the size of traded volume. Indeed, the excess kurtosis of the swapped amount of token Y, while still positive, is up to two orders of magnitude smaller than that of the returns.

Next, we analyze the decay of the excess kurtosis viewed as a function $K(n)$ of the number of aggregated swaps (or blocks) $n$, with $n$ smaller than or equal to 5. Table \ref{tab:kurt_hf} shows the result of the regression $K(n) \sim A \cdot n^{-p}$ (estimated by regressing the $\log(K(n))$ against $\log(n)$).

\begin{table}[h]
{\footnotesize
\centering
\begin{tabular}{lllll|lrrrr}
\toprule
\textbf{Pool} & \textbf{N=2}   & \textbf{N=3}   & \textbf{N=4}   & \textbf{N=5}   & \textbf{Pool} & \textbf{N=2}    & \textbf{N=3}   & \textbf{N=4}   & \textbf{N=5}   \\
\midrule
\multirow{2}{*}{\texttt{USDC-WETH} 0.01\%} & \textbf{0.928} & \textbf{0.752} & 0.443 & 0.671 & \multirow{2}{*}{\texttt{WBTC-WETH} 0.3\%}    & \textbf{1.200}  & \textbf{1.246} & \textbf{1.350} & \textbf{1.628} \\
 & \textbf{0.773} & 0.437 & \underline{0.478} & 0.856 & & \textbf{0.916}  & \underline{1.919}    & \underline{1.461} & \underline{1.961} \\
\multirow{2}{*}{\texttt{USDC-WETH} 0.05\%} & \textbf{1.039} & \textbf{0.945} & \textbf{0.983} & \textbf{1.104} & \multirow{2}{*}{\texttt{LINK-WETH} 0.3\%}    & \textbf{0.975}  & \textbf{1.078} & \textbf{1.085} & \textbf{1.264} \\
 & \textbf{2.370} & 0.785 & 0.935 & 0.961 & & \textbf{1.331}  & \textbf{1.504} & \textbf{1.610$^*$} & \textbf{1.683$^*$} \\
\multirow{2}{*}{\texttt{USDC-WETH} 0.3\%}  & \textbf{1.648} & \textbf{1.798} & \textbf{1.710$^*$} & \textbf{1.781$^*$} & \multirow{2}{*}{\texttt{MNT-WETH} 0.3\%} & \textbf{1.035}  & \textbf{1.023} & \textbf{1.390} & \textbf{1.522} \\
 & \textbf{2.477} & \textbf{2.269} & \textbf{2.127$^*$} & \textbf{1.790} & & \textbf{1.336}  & \textbf{1.450} & \textbf{1.423$^*$} & \textbf{1.325$^*$} \\
\multirow{2}{*}{\texttt{WETH-USDT} 0.01\%} & \textbf{0.915} & \textbf{0.853} & \textbf{0.782} & 0.555 & \multirow{2}{*}{\texttt{UNI-WETH} 0.3\%} & \textbf{1.911}  & \textbf{2.021} & \textbf{1.780} & \textbf{1.482} \\
 & \textbf{2.609} & 0.576 & 0.993 & 0.655 & & \textbf{1.543}  & \textbf{1.755} & 1.250 & 1.166 \\ \cline{6-10} 
\multirow{2}{*}{\texttt{WETH-USDT} 0.05\%} & \textbf{0.580} & \textbf{0.916} & 0.433 & 1.283 & \multirow{2}{*}{\texttt{USDC-USDT} 0.01\%}   & \textbf{0.563}  & 2.185 & 1.106 & 0.658 \\
 & \textbf{1.521} & \textbf{1.851} & \textbf{1.648} & \textbf{1.571$^*$} & & \textbf{0.393}  & 0.182 & 0.151 & 0.061 \\
\multirow{2}{*}{\texttt{WETH-USDT} 0.3\%}  & \textbf{0.967} & \textbf{1.189} & \textbf{1.210} & \textbf{1.345} & \multirow{2}{*}{\texttt{USDC-USDT} 0.05\%}   & \textbf{0.664}  & \underline{0.392}    & \underline{0.492} & \underline{0.613} \\
 & \textbf{1.131} & \underline{2.097} & \underline{1.807} & \underline{1.497} & & \textbf{0.501}  & \textbf{0.397} & 0.835 & 0.734 \\
\multirow{2}{*}{\texttt{WBTC-USDC} 0.05\%} & \textbf{0.802} & \textbf{0.714} & 0.424 & 0.259 & \multirow{2}{*}{\texttt{DAI-USDC} 0.01\%}    & \textbf{5.097}  & 1.437 & 2.007 & 1.803 \\
 & \textbf{1.833} & \textbf{1.341} & \textbf{1.420} & 1.016 & & \textbf{-0.107} & \textbf{0.087} & \underline{0.006} & \underline{0.263} \\
\multirow{2}{*}{\texttt{WBTC-USDC} 0.3\%}  & \textbf{1.179} & \textbf{1.660} & \textbf{1.737} & \textbf{1.818$^*$} & \multirow{2}{*}{\texttt{DAI-USDT} 0.01\%}    & \textbf{0.960}  & 0.673 & \textbf{0.734} & 0.468 \\
 & \textbf{1.406} & \textbf{1.435$^*$} & \textbf{1.469$^*$} & \textbf{1.766} & & \textbf{-0.567} & 0.502 & 0.393 & 0.739 \\
\multirow{2}{*}{\texttt{WBTC-USDT} 0.05\%} & \textbf{0.533} & \textbf{0.494} & 0.124 & 0.152 & \multirow{2}{*}{\texttt{USDe-USDT} 0.01\%}   & \textbf{0.999}  & 0.567 & \underline{0.838} & \underline{0.936} \\
 & \textbf{0.148} & 0.776 & 0.603 & \underline{0.786} & & \textbf{0.446}  & \textbf{0.740} & 1.078 & 0.996 \\ \cline{6-10} 
\multirow{2}{*}{\texttt{WBTC-USDT} 0.3\%}  & \textbf{0.758} & \underline{2.150} & 1.385 & 1.272 & \multirow{2}{*}{\texttt{WETH-weETH} 0.01\%}  & \textbf{0.533}  & \underline{1.130}    & 0.758 & 0.681 \\
 & \textbf{1.948} & \textbf{2.000$^*$} & \textbf{1.929$^*$} & \underline{1.529} & & \textbf{1.748}  & \textbf{1.512} & \underline{1.084} & \underline{0.939} \\ \cline{1-5}
\multirow{2}{*}{\texttt{WBTC-WETH} 0.01\%} & \textbf{0.545} & \textbf{0.777} & \textbf{0.701} & 0.427 & \multirow{2}{*}{\texttt{wstETH-WETH} 0.01\%} & \textbf{0.772}  & 0.233 & 0.249 & 0.267 \\
 & \textbf{3.194} & \textbf{2.546} & 1.712 & 1.460 & & \textbf{0.996}  & \textbf{0.934} & \textbf{0.953} & \underline{1.253} \\
\multirow{2}{*}{\texttt{WBTC-WETH} 0.05\%} & \textbf{2.666} & 0.650 & 2.845 & 1.474 & \multirow{2}{*}{\texttt{WBTC-LBTC} 0.05\%}   & \textbf{0.804}  & \textbf{0.678} & \textbf{0.832} & \textbf{0.762} \\
 & \textbf{1.092} & \textbf{1.228} & \textbf{1.094} & \textbf{1.167} & & \textbf{1.149}  & \textbf{1.096} & \textbf{1.089} & \textbf{1.116$^*$} \\
\bottomrule
\end{tabular}
\caption{Excess kurtosis as a function of the number of aggregated swaps (top row) and blocks (bottom row). The $p$ coefficient from the regression $K(n) \sim A\cdot n^{-p}$ is shown. The regression is performed using up to $N$ values, and every column is for $N$. Coefficients obtained from a regression with the $R2$ score greater than $0.9$ ($0.75$) are in bold (underlined). $^*$ highlights those coefficients whose $95\%$ confidence interval (according to the t-test) is on the right of $1$.}
\label{tab:kurt_hf}
}
\end{table}

We find that some pools (especially those with a high fee tier) have a decay coefficient $p$ greater than or equal to $1$. This pattern, which is theoretically associated with short memory in the stochastic variance process, represents a deviation from what is typically observed in TradFi. Contrastingly, when working with larger aggregation values $n$, we expect to recover a slow-decaying kurtosis function. Table \ref{tab:kurt} in Appendix \ref{app:seqr} shows the results of the power-law fit and the mesokurtosis test proposed in \cite{anscombe1983distribution} with the null hypothesis of a mesokurtotic distribution. The number of aggregated swaps $n$ ranges from $15$ to the number of swaps in the pool divided by $20$, as $20$ is the minimum sample size for performing the mesokurtosis test. The power-law provides an acceptable fit in almost all the pools. Generally, the estimated coefficient $p$ decreases as the fee tier $\phi$ increases, and they lie in the range $[0.2, 1]$.

%
\subsection{Returns Autocorrelation}
\label{subsec:ret_acf}
\begin{figure}[h]
	\centering
	\includegraphics[width=\linewidth]{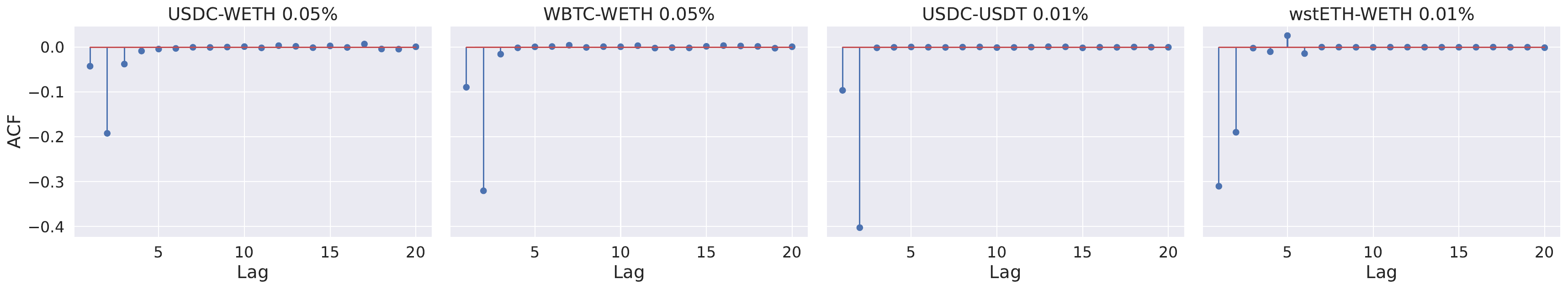}
	\caption{Returns ACF in swap-time, up to lag 20. The red area corresponds to the significance level, which is obtained with a bootstrap approach. Each subplot is for a pools cluster: \texttt{USDC-WETH} 0.05\% (Normal cluster); \texttt{WBTC-WETH} 0.05\% (Volatile cluster); \texttt{USDC-USDT} 0.01\% (Stable cluster); \texttt{wstETH-WETH} 0.01\% (Synthetic cluster).}
	\label{img:acf_event}
\end{figure}

Another set of quantities to investigate is the returns AutoCorrelation Function (ACF). Figure \ref{img:acf_event} shows the ACF, in swap-time, of the most traded pool in each cluster. The significance region is estimated using a bootstrap procedure: we generate 500 i.i.d. bootstrap time series from the original data and compute the ACF for each. For every lag, the bounds of the significance region are given by the $0.05$ and $0.95$ quantiles of the bootstrapped ACF distribution.

Interestingly, we find strong negative and significant autocorrelation for the first $3$ lags. The magnitude of the peaks depends on the type and fee rate of the pool. For example, for Normal and Volatile clusters, this pattern is mainly present in low fee pools and is more pronounced at the second lag, while for Synthetic pools, we also observe large negative autocorrelation at lag $1$. Finally, we find that the second lag peak is persistent when computing the ACF in tick-time, however, it vanishes when working in clock-time. 

\begin{figure}[H]
	\centering
	\includegraphics[width=\linewidth]{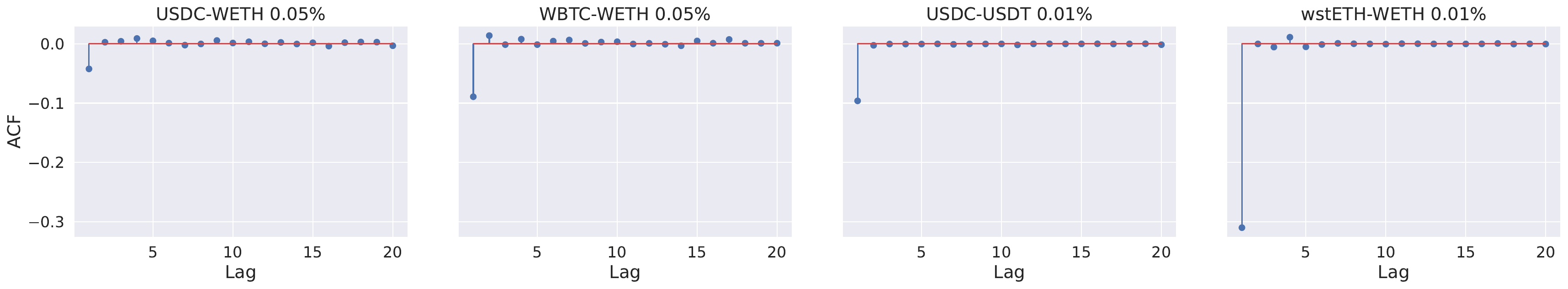}
	\caption{Returns ACF in swap-time, up to lag $20$, conditional on the absence of sandwich attacks between two lags. The red area corresponds to the significance level, which is obtained with a bootstrap approach. Each subplot is for a pools cluster.}
	\label{img:acf_conditioned_no_sandwich}
\end{figure}

To understand the factors driving this pattern, we investigate each lag separately. First, let's focus on the peaks at lag $2$ and $3$. In this case, we conjecture that they are due to sandwich attacks discussed in Subsection \ref{subsec:sand_att}. This is consistent with the observation that the pattern typically appears in pools with low fees, where front-running is more profitable. Moreover, it is coherent with the persistence of the ACF in tick-time and its vanishing in clock-time. Indeed, front-run and back-run trades are of significant volume and, consequently, they typically move the price tick. Furthermore, all the swaps involved within a single sandwich are in the same block, thus they share the same timestamp. To investigate our claim, we compute the ACF of log returns, conditioning on the absence of sandwich attacks between time $t$ and  $t+l$ ($l > 0$, swap-time). Figure \ref{img:acf_conditioned_no_sandwich} shows these results for the same pools as in Figure \ref{img:acf_event}. When doing so, we find that the strong negative peaks at lags $2$ and $3$ we observe in the unconditional ACF disappear, while the autocorrelation at the first lag, due to a different strategy, survives.

Next, we focus on the negative autocorrelation at lag $1$. We recall that in TradFi, negative autocorrelation at lag $1$ is due mainly to the bid-ask bounce effect at high frequencies. In DEXs, however, there is no spread, and therefore, one should not expect a bid-ask bounce. We, instead, conjecture that the negative autocorrelation at lag $1$ is driven by the presence of echo swaps and the action of \textit{reverse arbitrageurs}\footnote{
Conversely, there exist stale-price arbitrageurs who profit by transmitting the impact of new information -- i.e., changes in the fundamental value of the tokens -- from centralized exchanges (CEXs) to decentralized exchanges (DEXs).}, that is, a class of arbitrageurs who trade against the price impact generated by noise traders \cite{capponi2025liquidity}. As for the echo swaps, we find that the ACF in the absence of their contribution is larger than that when including all the swaps in $21$ out of $23$ pools with negative first-lag ACF. This result is insensitive to whether we include or exclude sandwich attacks from the ACF computation. Regarding the reverse arbitrageurs,

findings in \cite{capponi2025liquidity} highlight that this strategy is profitable if the impact is larger than the pool fee rate $\phi$\footnote{To be completely accurate, we should also consider the gas fee. Our aim here, however, is to provide the intuition for the negative autocorrelation observed at the first lag. Thus, we keep the framework as simple as possible.} To test the relevance of reverse arbitrageurs, let $\epsilon_t$ be the sign of the swap occurring in swap-time $t$, $r_t$ be the log return between $t$ and $t-1$, and $\phi$ be the fee tier of the pool, we then compute the conditional probabilities $p_{g} = \mathbb{P}\big[\epsilon_{t+1} \neq \epsilon_t \ \big| \ |r_t| > \phi\big]$ and $p_{l} = \mathbb{P}\big[\epsilon_{t+1} \neq \epsilon_t \ \big| \ |r_t| \le \phi\big]$. That is, we compute the probability of a swap sign reversal, after a swap at time $t$, conditional on the absolute return exceeding/falling short of $\phi$. We then perform a one-sided bootstrap test with $1,000$ replicas under the null hypothesis $H_0:p_g \le p_l$. The test produces a p-value smaller than $0.01$ across all pools; hence, we reject $H_0$. Further evidence for our conjecture is provided by computing $Corr(r_{t+1}, r_t)$ conditional on $|r_t| > \phi$ and $|r_t| \le \phi$. Conditional on $|r_t| > \phi$, we recover the negative correlation observed. Conditional on $|r_t| \le \phi$, the ACF shows a slightly positive trend that is explained by the presence of herd effect between LT, as we discuss in Section \ref{subsec:trade_sign}. Further quantitative results are reported in Appendix \ref{app:seqr}, Table \ref{tab:l1d}.

\subsection{Returns Magnitude Autocorrelation}

\begin{figure}[h]
	\centering
	\includegraphics[width=\linewidth]{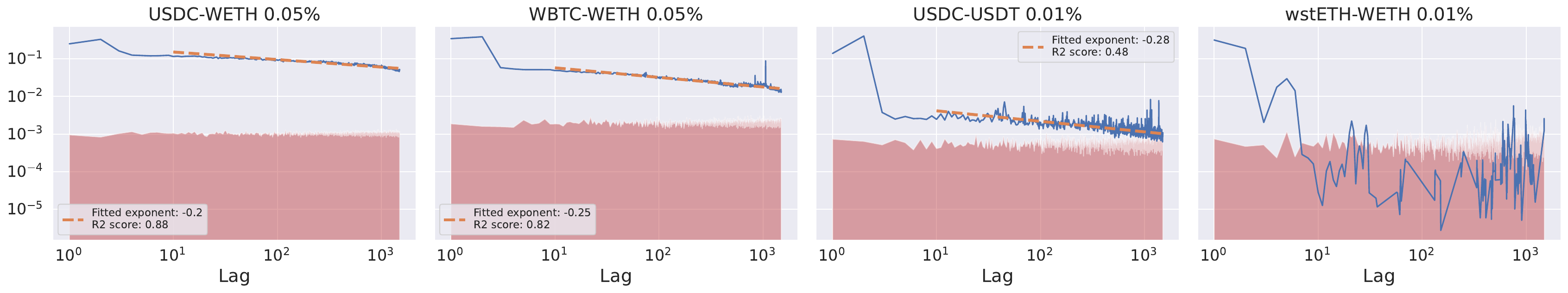}
	\caption{Autocorrelation in the absolute returns, swap-time. The panels show the most traded pool in each cluster. For sufficiently large lags, the data is fit well by a power-law $ACF(L) \sim A\; L^{p}$. The exponent $p$ and the $R2$ score are reported in the legend. The significance level described by the red area is obtained via the bootstrap approach described in Subsection \ref{subsec:ret_acf}.}
	\label{img:acf_absret}
\end{figure}

In this section, we investigate whether these markets exhibit volatility clustering. We do so by studying the autocorrelation of the absolute value of returns. Figure \ref{img:acf_absret} shows the ACF in swap-time for the most traded pool in each cluster. 

In almost all the Normal and Volatile pools, we find a decay consistent with long-memory. Contrastingly, we find that Stable and Synthetic pools have ACFs that are consistent with short-memory, except for \texttt{USDC-USDT} 0.01\%. The fitted coefficients of the power-law $ACF(L) \sim A \cdot L^p$, pool-by-pool, are provided in Appendix \ref{app:seqr}, Table \ref{tab:lm_rrts}. The exponent $p$ typically increases with the fee tier $\phi$ and is in the interval $[0.2, 0.5]$ for almost all the pools, hence, the exponent $p$ is roughly in line with the literature on TradFi \cite{chakraborti2011econophysics}. We note that these results are robust to switching to both tick-time and clock-time. As expected, the decay becomes faster as we reduce the observation frequency.
\subsection{Volume \& Signed Volume Autocorrelation}
\label{subsec:vol_sign}
\begin{figure}[h]
	\centering
	\includegraphics[width=\textwidth]{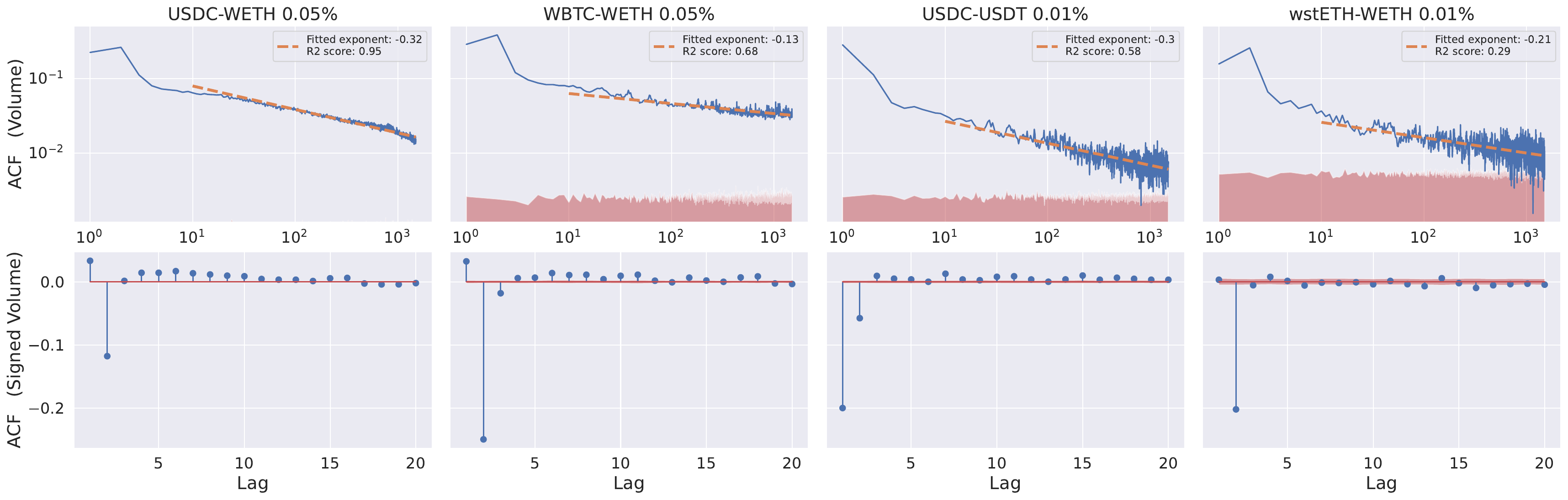}
	\caption{Volume ACF (upper) and signed volume ACF (lower) in swap-time. The panels show the most traded pool in each cluster. The legend reports the coefficient and the $R2$ score of the power-law fit. The significance level described by the red area is obtained via the bootstrap approach described in Subsection \ref{subsec:ret_acf}.}
	\label{img:acf_vol_line_x2}
\end{figure}

We now turn to the analysis of volume and signed volume. The volume is computed as the unsigned numeraire volume exchanged. In swap-time, this is simply the volume of the swap, whereas at aggregated timescales we use the sum of volumes in a given time interval. Similarly, the signed volume time series is obtained by multiplying the numeraire volume by the sign $\epsilon$, which equals $+1$ if the swap occurs from the first to the second token of the pool, otherwise equals $-1$. Figure \ref{img:acf_vol_line_x2} shows the volume and signed volume ACFs for the most traded pool in each cluster. The observed decays are consistent with long-memory in the volume ACF for almost all the pools, with only a few exceptions among Stable and Synthetic pools. See Appendix \ref{app:seqr}, Table \ref{tab:lm_rrts} for a comprehensive summary of the power-law fitted coefficients. Contrastingly, the signed volume is short-memory and shows the same negative peaks at the first lags as the ACF of returns. These findings are consistent with the price formation rule in Uniswap v3. As for the returns, the explanation behind such a pattern is related to the activity of MEV searchers performing sandwich attacks. When we remove events related to the activity of MEV searchers, the resulting ACF is similar to those shown in Figure \ref{img:acf_conditioned_no_sandwich}.

\begin{figure}[h]
	\centering
	\includegraphics[width=\textwidth]{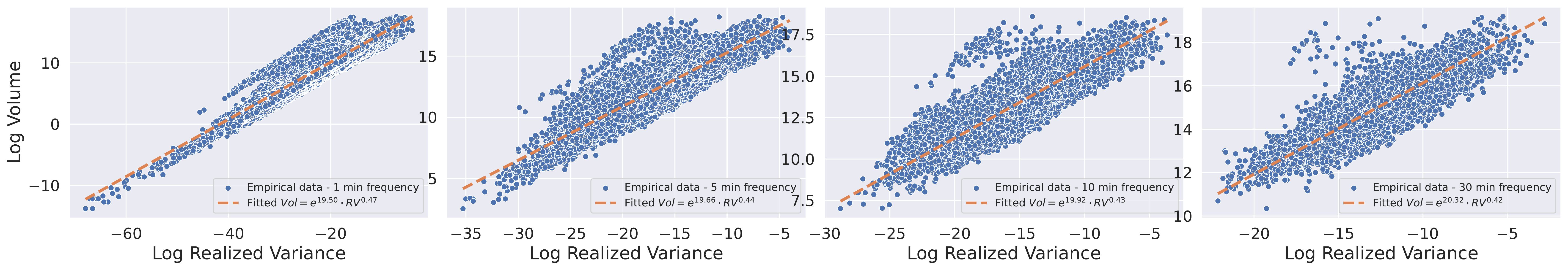}
	\caption{\texttt{USDC-WETH} 0.05\% pool - Swaps Volume (referred to USDC) vs Realized Variance (RV), both on a log-scale. They are synchronously computed on time windows of fixed length. The orange dashed line shows the linear fit between the logarithms.}
	\label{img:stf_vol_vs_rv_min_usdc_weth_005_line}
\end{figure}

Finally, we study the relationship between volume and variance by comparing the logarithms of Realized Variance (RV) and Volume. To this end, we partition the dataset according to clock-time at different aggregations. Inside each time interval, we compute the RV and the aggregated volume, and we investigate the contemporaneous relation between these quantities. The result concerning the \texttt{USDC-WETH} 0.05\% pool is shown in Figure \ref{img:stf_vol_vs_rv_min_usdc_weth_005_line}.

As shown, we find almost a linear relationship between the logarithms; moreover, the fitted coefficients are roughly the same at each frequency. Hence,  the RV is approximately a power of the volume. This result holds for almost all the pools. Table \ref{tab:vol_vol} in Appendix \ref{app:seqr} gives further insights into the linear relationship between $\log(RV)$ and $\log(Vol)$.
\subsection{Trade Direction Autocorrelation}
\label{subsec:trade_sign}
\begin{figure}[h]
	\centering
	\includegraphics[width=\textwidth]{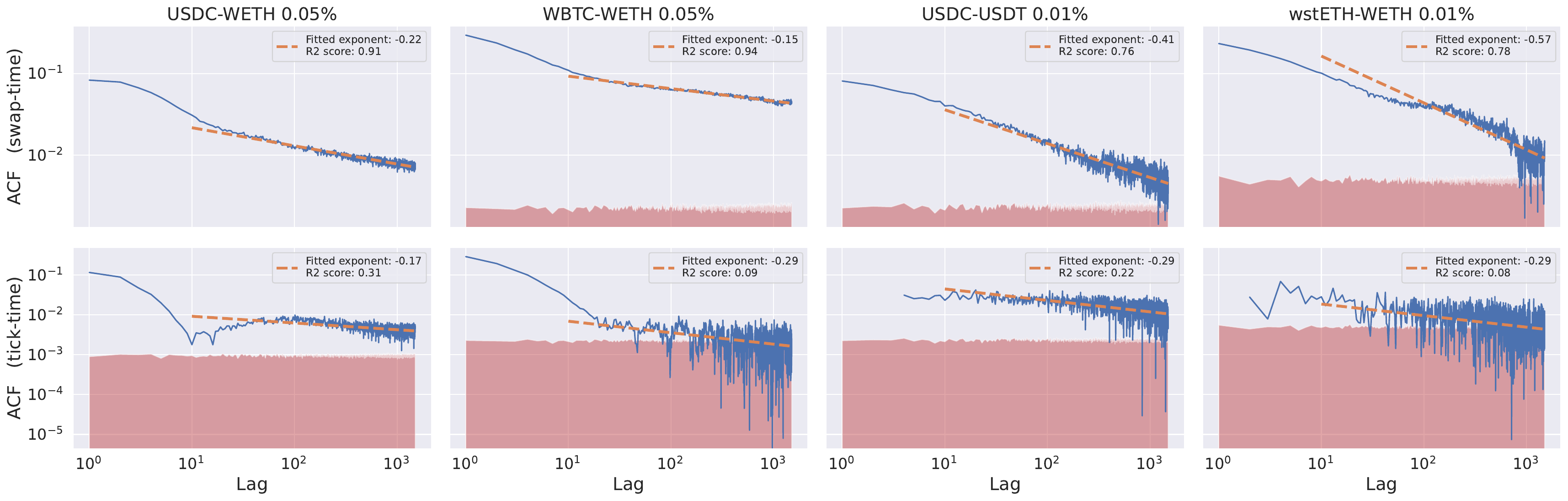}
	\caption{Sign ACF in swap-time (upper) and tick-time (lower). The panels show the most traded pool in each cluster. The output of the power-law fit is in the legend. The significance level is obtained via the bootstrap approach described in Subsection \ref{subsec:ret_acf}.}
	\label{img:acf_sign_line_x2}
\end{figure}

An important feature often observed in TradFi is the slow decay of the trade signs time series ACF (see, for example, \cite{lillo2004long} and \cite{bouchaud2009markets}), both in event-time and tick-time \cite{chakraborti2011econophysics}. Here, we investigate whether this pattern survives in DEXs. As discussed in the previous subsections, the sign of a swap is defined as $+1$ or $-1$, depending on whether the swap is from the first to the second token or the reverse. Thus, the sign series in swap-time is a binary sequence of $\pm1$. In tick time, however, we define the sign as the mean of the signs of all the trades between two consecutive tick changes. Figure \ref{img:acf_sign_line_x2} shows the ACF of the sign time series for the most traded pool in each cluster in swap-time and tick-time.

Interestingly, we observe a slow decay of the trade signs ACF also in our data. This phenomenon is particularly evident for Stable and Synthetic pools, while it is less pronounced in Normal and Volatile ones. The full picture is in Appendix \ref{app:seqr}, Table \ref{tab:lm_rrts}. To investigate the causes that originate the slow decaying observed, we draw from the TradFi literature, where mainly two explanations have been proposed. The most respected one attributes the long-memory property to the execution of meta-orders\footnote{A meta-order is a large order split into multiple smaller ones executed incrementally to avoid incurring huge transaction costs.}. It is not clear, however, whether this strategy is equally relevant for DEXs. On the one hand, it can be helpful to mitigate sandwich attacks and exploit mean-reversion to reduce execution costs; on the other hand, every swap generates an additional gas fee, which raises doubts about the meta-order profitability. The second explanation proposed in TradFi relates the slow decline to the herding effect -- that is, the tendency of investors to replicate the trades of others \cite{lebaron_yamamoto}.

This hypothesis has been found to be less compelling in traditional markets. This hypothesis may, however, be more relevant in DEXs given their underlying dynamics. Indeed, the cryptocurrency price is characterized by many strong price trends that can be attributed to herding effects. Such patterns introduce strong autocorrelation in the direction of the swaps.

\begin{figure}[h]
    \centering
    \includegraphics[width=\textwidth]{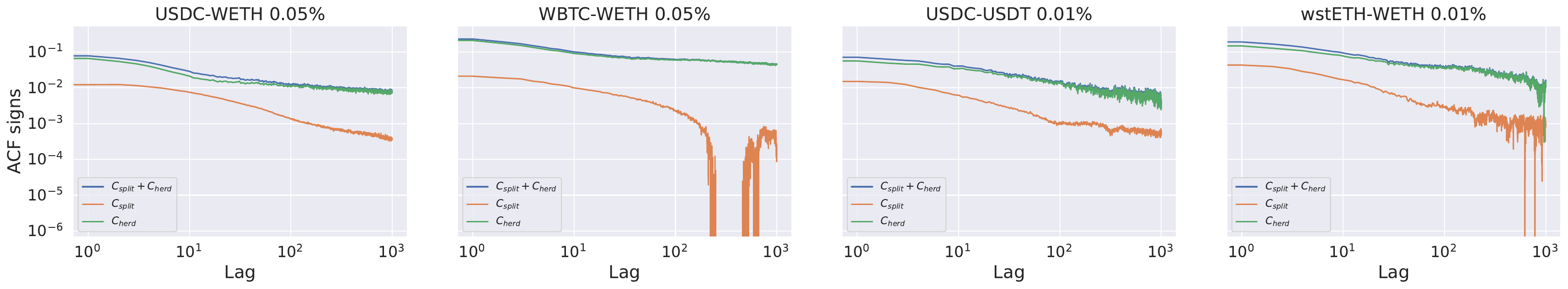}
    \caption{Decomposition of the ACF of swap signs (swap-time) in two components: one for the swaps performed by the same wallet ($C_{split}$) and one for the swaps performed by different wallets ($C_{herd}$). Their sum is equal to what is shown in the top panels in Figure \ref{img:acf_sign_line_x2}.}
    \label{img:acf_sign_split_vs_herd}
\end{figure}

We empirically compare the reliability of the two explanations by following the procedure described in \cite{toth}. Specifically, we decompose the autocorrelation function in order to disentangle, at each lag, the contribution of the swaps performed by the same wallet ($C_{split}$) from the swaps performed by different wallets ($C_{herd}$). Specifically, let $C(\tau)$ be the ACF evaluated at lag $\tau$ and $\epsilon_t^i$ the action of wallet $i$ at time $t$, defined as +1 for a swap from the first to the second token, $-1$ for the opposite direction, and $0$ if the swap is not performed by wallet $i$. Specifically, we define

\begin{equation}
    C(\tau) :=  \frac{1}{N} \sum_t \sum_{i,j} \epsilon_t^i \epsilon_{t+\tau}^j - \left(\frac{1}{N} \sum_t \sum_i \epsilon_t^i\right)^2\,.
\end{equation}
This expression may be rearranged as
\begin{equation}
 C(\tau)  = C_{split}(\tau) + C_{herd}(\tau)
\end{equation}
where
\begin{align}
    C_{split}(\tau) &:= \frac{1}{N} \sum_i \left[\sum_t \epsilon_t^i \epsilon_{t+\tau}^i - \frac{1}{N} \left(\sum_t \epsilon^i_t\right)^2\right] 
    \\ 
    C_{herd}(\tau) &:= \frac{1}{N} \sum_{i\neq j} \left[\sum_t \epsilon_t^i \epsilon_{t+\tau}^j - \frac{1}{N} \left(\sum_t \epsilon_t^i\right)\left(\sum_t \epsilon_t^j\right)\right]
\end{align}
The results of this decomposition are reported in Figure \ref{img:acf_sign_split_vs_herd}, where we observe that the herding component dominates the autocorrelation values. This is a notable departure from trends seen in traditional markets.

It is worth noting that this result is also observed when accounting for some specific features of Uniswap. First, the result is qualitatively the same when (pure and mixed) sandwich swaps are removed. Second, the result is robust to \textit{routing}: when users want to submit an order on Uniswap, they can either interact directly with specific pools or rely on Uniswap’s routing system (or third-party protocols like \texttt{1Inch} or \texttt{CoW} Swap). The former method is preferable for performing the MEV strategies discussed in the previous sections, whereas the latter leverages specialized router smart contracts to identify the most efficient way to execute the trade, in order to achieve the best possible execution price. Specifically, router smart contracts often trade the same pair on pools with different fee tiers\footnote{For large swaps, the impact generated on the most favorable pool could make it profitable to execute a portion of the swap on the other pools.} or split the swap into smaller parts and route them across multiple pools (with different pair) and/or multiple DEXs. Therefore, what we find in a specific pool can be part of a path originating from the intention to swap between two other tokens\footnote{For example, if the market condition is favorable, a trade from USDC to WBTC can be split into: trade from USDC to USDT; trade from USDT to WETH; trade from WETH to WBTC.}.

\begin{figure}
    \centering
    \includegraphics[width=0.7\linewidth]{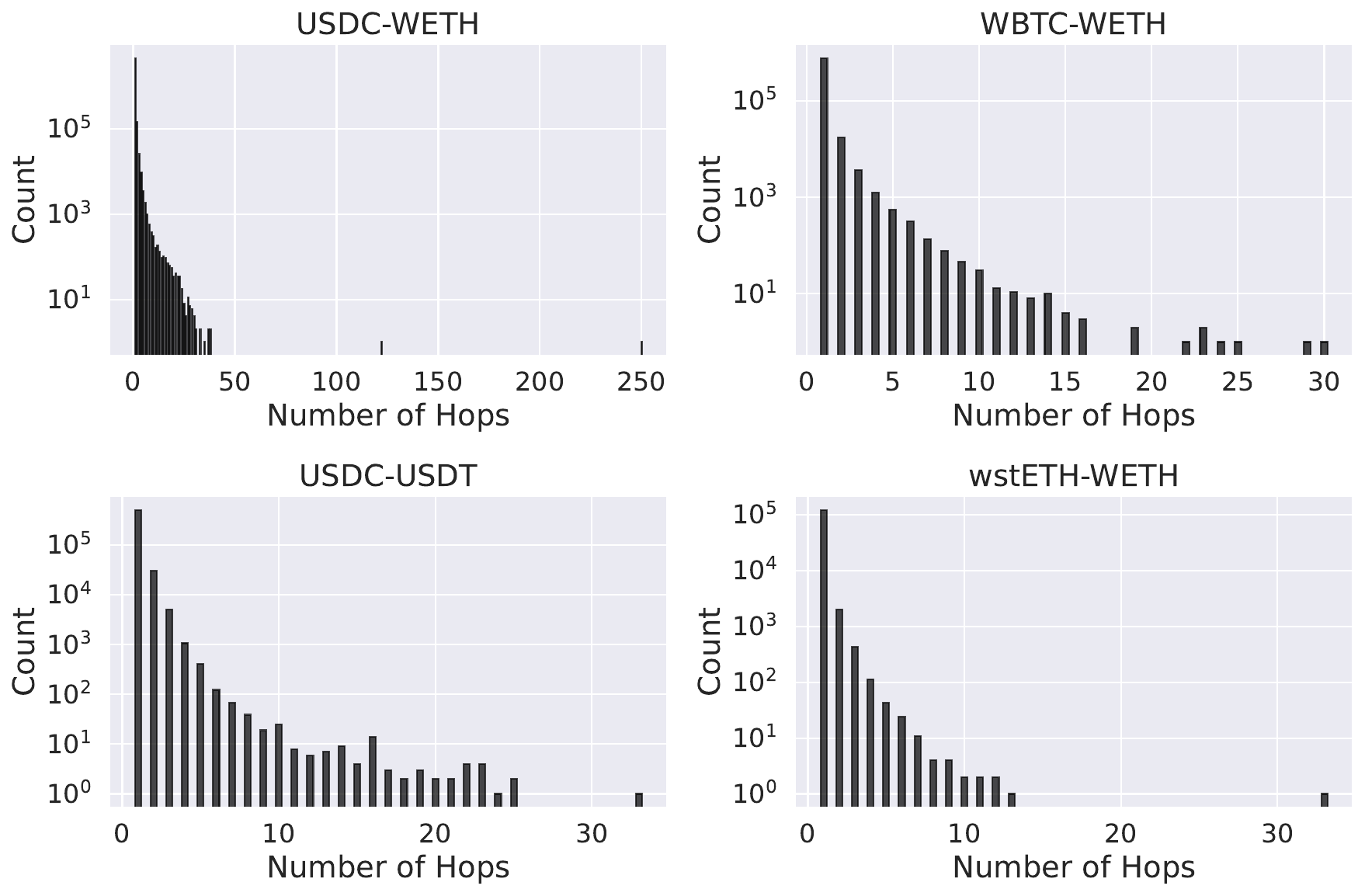}
    \caption{Observed distribution of routing path lengths (in hops) between selected token pairs during routing.}
    \label{img:n_hops}
\end{figure}

One may ask whether the sign persistence previously observed at the ``pools'' level survives at the ``routing'' level -- i.e., when accounting for these dynamics. To assess the cause, we track token flows associated with each transaction hash\footnote{Usually, a smart routing swap is atomic. Therefore, all the transactions share the same hash.}, clearly identifying both the input and output tokens along with the actual direction of the aggregated swap event. In this way, we track the actual `intention' of the user's swap. Figure \ref{img:n_hops} shows the number of steps to go from one token to the other for the most active pair in each group. We then apply exactly the same analytical framework employed at the single-pool level. Ultimately, the results are consistent. 

%
%
%
\section{Statistics of the Liquidity}
\label{sec:liquidity}
Liquidity provision is the most important source of revenue for investors in Uniswap -- see \cite{risk2025dynamics} for a detailed analysis of liquidity at medium-frequency. An LP can add tokens to the pool and earn fees whenever a swap occurs. Therefore, it is important to study the statistical properties of liquidity in pools to acquire insights about how investors choose to act. In this section, we focus our attention on the liquidity minted, as the liquidity burned shows similar patterns. When providing liquidity, the two key quantities of interest are the amount minted and the width of the provision range. We compute their ACF to study the temporal persistence. For the liquidity minted, we separately study active and passive liquidity. Regarding the width range, we only consider non-JIT events, as JITs range is almost always set to the pool's \texttt{tickSpacing} parameter. Generally, we do not find ACF consistent with long-memory. The only exception is the liquidity minted by JIT attackers. Indeed, in the pools where JIT occurrences are sufficient (more than $2,000$), we usually find a slow decay.

\begin{figure}[h]
	\centering
	\includegraphics[width=\linewidth]{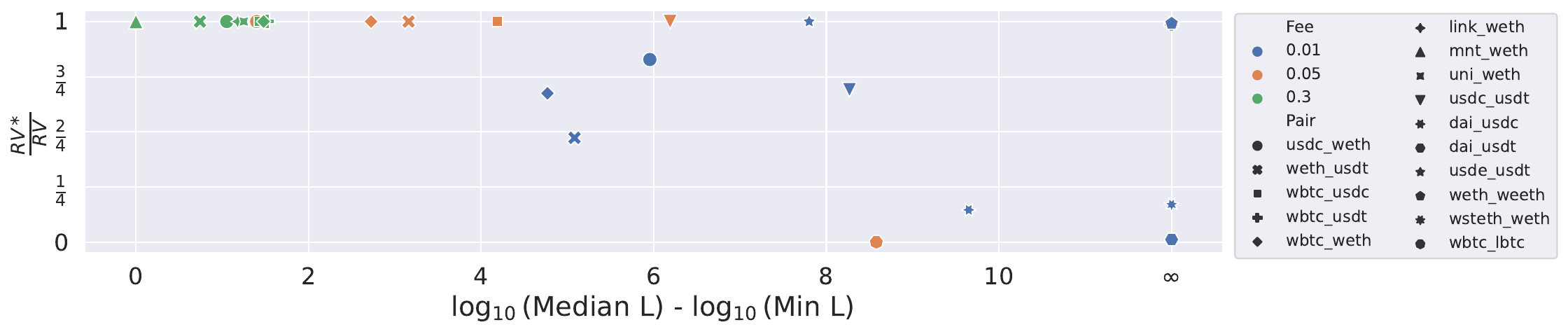}
	\caption{Differences in the Realized Variance (RV) as a function of the strength of liquidity droughts, measured as the difference between the median and minimum log liquidity. $RV$ is for the annualized Realized Variance, computed from marginal prices sampled every 10 minutes over the entire period. $RV*$ is defined analogously, but only includes swaps whose corresponding active liquidity is above the 5\% quantile of the pool's active liquidity. The infinite value on the x-axis is for those pools whose minimum active liquidity is 0.}
	\label{img:delta_rv_L}
\end{figure}

An interesting pattern observed in DEXs is related to the liquidity droughts. That is, sometimes the active liquidity suddenly drops close to $0$. These drops are triggered by swaps rather than burn events, so they are due to the Concentrated Liquidity mechanism. As a reference, one can consider Table \ref{tab:SrL_summary} in Appendix \ref{app:oep}, where the minimum active liquidity in the considered time range (pool-by-pool) is compared to the median one. The Table shows that four pools reach a minimum equal to $0$, and another four display a minimum not exceeding $100$. Generally, large liquidity drops may lead to a significant amplification of the price impact, resulting in abnormal absolute returns and deviations from the CEX benchmark, which in turn lead to an ``artificial'' increase in the volatility measures (such as the RV), as shown in Figure \ref{img:delta_rv_L}.

\begin{figure}[h]
	\centering
	\includegraphics[width=\linewidth]{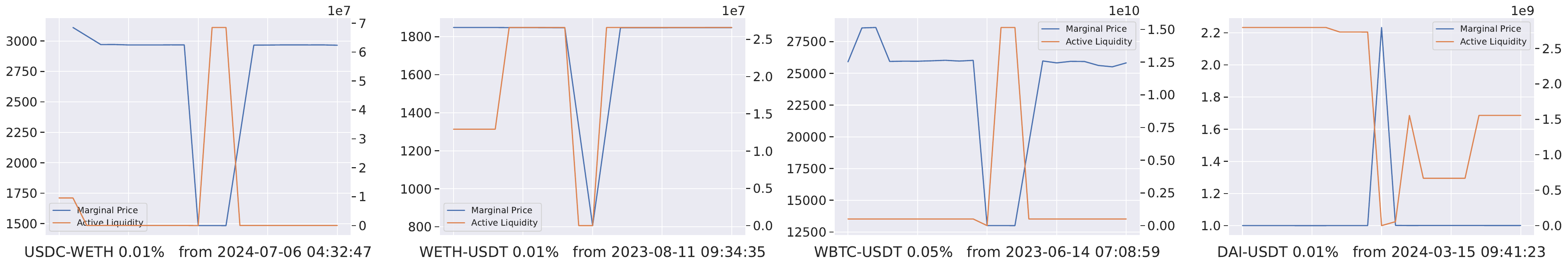}
	\caption{Liquidity droughts and corresponding price peaks. Each subplot reports one example covering 10 events before and after the liquidity drop. The blue line is for the marginal price time series; the orange line is for the active liquidity. The x-label specifies the pool and the date of the first event of the example. The x-axis is in event-time. Two y-axes for each subplot: the left one is for the price scale; the right one is for the liquidity scale.}
	\label{img:example_peaks_droughts}
\end{figure}

However, these deviations from the CEX price are not profitable. Indeed, due to liquidity droughts, any trade would cause a significant impact, which immediately realigns the price with the benchmark and draws the liquidity into the normal regime.
Finally, Figure \ref{img:example_peaks_droughts} illustrates an example of the effect of liquidity droughts, which result in spikes in the marginal price time series.
%

%
\subsection{Provision Range vs Volatility}
\label{subsec:range_vs_vol}

\begin{figure}[h]
	\centering
	\includegraphics[width=\linewidth]{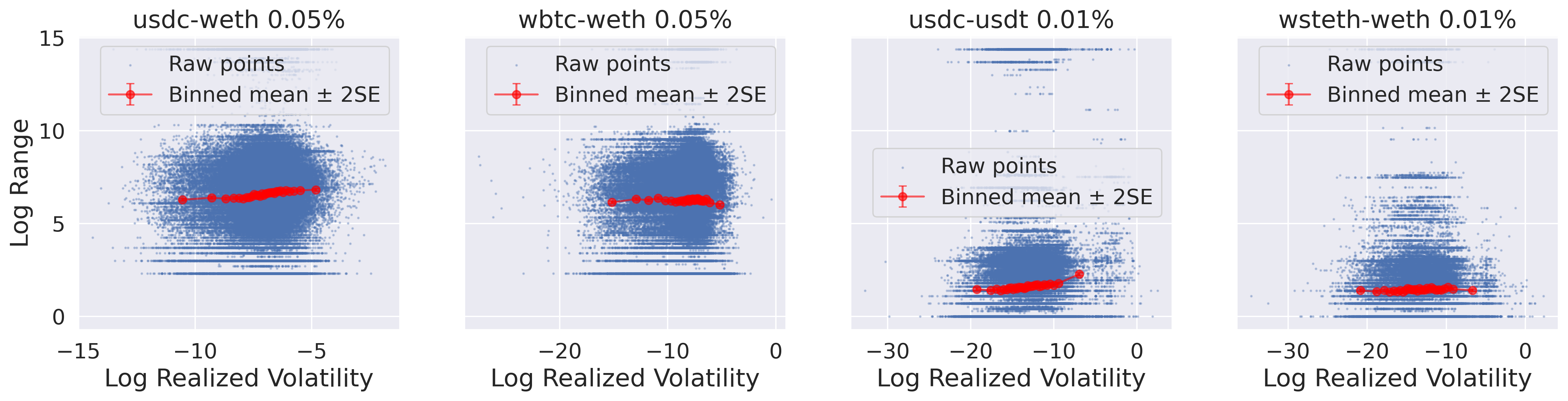}
	\caption{Liquidity provision width (Range) vs RV for the most active pools for each group. The RV is computed over a time window of $10$ minutes before the mint. The dotted red line represents the average log provision width together with the standard errors calculated within each of the $30$ equal sized bins into which the log RV values have been divided.}
	\label{img:width_vs_vol}
\end{figure}

We next investigate the relationship between the liquidity provision range and the volatility. The intuition based on \cite{fan2022differential}, \cite{cartea2024decentralized}, and \cite{capponi2025liquidity} suggests that, with other market conditions fixed, the optimal (i.e., that which maximizes the expected utility) strategy for the LP is to widen the tick range when the volatility increases. Such reasoning applies only to passive liquidity, i.e., non-JIT mints. Therefore, we empirically study the relation between volatility and liquidity provision width. Figure \ref{img:width_vs_vol} shows an example related to the most traded pool in each cluster by plotting the logarithm of the range width against the logarithm of the Realized Variance, which we use as a proxy of the volatility. The RV is computed in a time window of ten minutes before the mint. Similar results are obtained for the other pools.

\begin{figure}[h]
	\centering
	\includegraphics[width=0.6\linewidth]{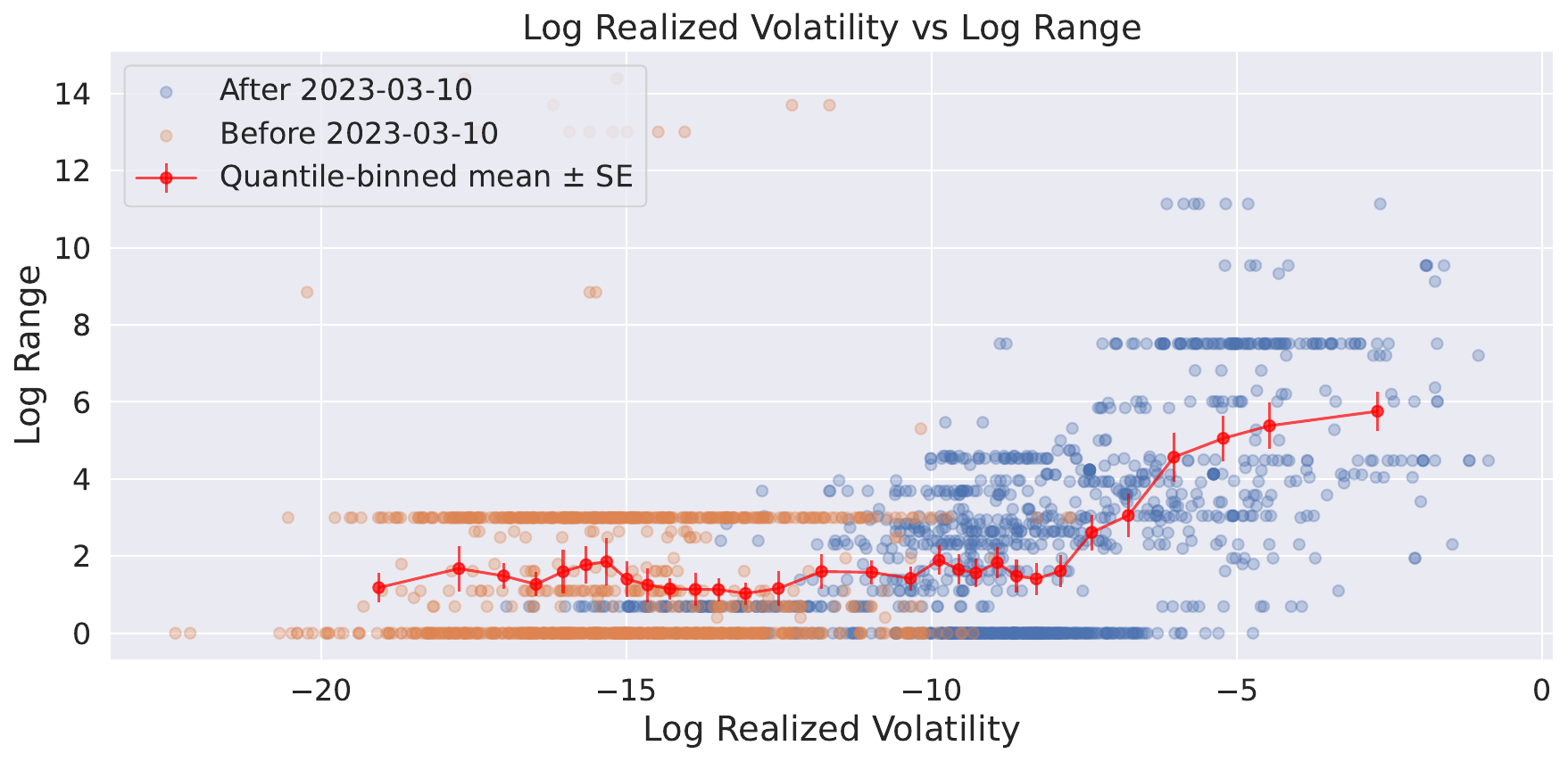}
	\caption{Liquidity Provision Width vs Realized Volatility (RV) for USDT-USDC 0.01\% from March 3, 2023  to March 17, 2023, i.e. 7 days before and after the SVB collapse. The RV is computed using a 10-minute time window preceding each mint event. The red points represent the average log provision widths calculated within each of the 30 equally sized bins into which the log RV values have been divided.}
	\label{img:svb_crisis_vol_vs_range}
\end{figure}

From the figure, the liquidity provision range is essentially independent of the volatility, resulting in a discrepancy between what we expect from theory and what we see in the markets. This can be due to several factors, such as irrational investors, different assumptions about the market behavior, and changes in the other market conditions (expected return, risk-aversion coefficient).  Other experiments with a longer time window ($30$ minutes) and with different proxies for the volatility (modulated RV, number of swaps, swapped volume) lead to the same conclusion. This evidence aligns with the findings of Capponi et al. \cite{capponi2025liquidity}, who demonstrated that liquidity providers (LPs) select wider price ranges not primarily due to increased exchange rate volatility itself, but rather due to increased risks of adverse selection from arbitrageurs. They leveraged the Silicon Valley Bank (SVB) crisis on March 10, 2023, as a natural experiment to test their model. This event caused a significant depeg of USDC. By comparing LP price ranges before and after the SVB event across Layer 1 (Ethereum) and Layer 2 platforms\footnote{A Layer 2 solution on Ethereum is a secondary framework built atop the main blockchain (Layer 1) that processes transactions off-chain to reduce congestion and lower gas fees. Layer 2 solutions improve scalability and transaction throughput while still benefiting from Ethereum’s robust security and decentralization} (Arbitrum and Optimism, which feature significantly lower gas fees for range adjustments), they effectively excluded the simpler hypothesis of an unconditional positive relationship between volatility and price range adjustments. Indeed, focusing specifically on the SVB crisis event, we observe an increase in the relationship between volatility and liquidity provision range, as illustrated in Figure \ref{img:svb_crisis_vol_vs_range}. This dynamics is not present in random periods of the same length taken from our data.

\begin{figure}[h]
    \centering
    \includegraphics[width=\linewidth]{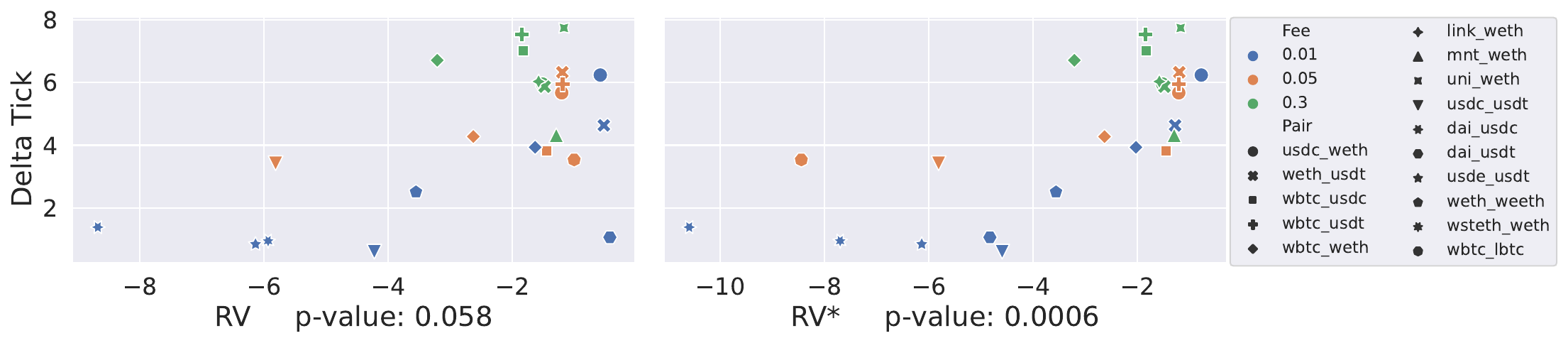}
    \caption{Provision's range versus Realized Variance, log-log scale. For every pool, the provision's range is expressed as the liquidity-weighted mean difference between the upper and lower ticks over all mint events. RV is for the Realized Variance (annualized) computed over the whole time series sampled at 10-minute frequency. RV* is computed in the same way, but removing observations corresponding to liquidity droughts (defined as active liquidity below the median $\times 10^{-4}$. Both axes are on a log scale. The p-value on the x-axis label refers to the Spearman correlation test with permutation.}
    \label{fig:ultimate_range_vs_rv}
\end{figure}

We carry out further analysis on the relationship between the provision's range width and the volatility through a cross-sectional study. For each pool, we compute the annualized RV from $10$-minute frequency, both with and without removing the peaks associated with liquidity droughts. Furthermore, we calculate the average difference between upper and lower ticks of the mint events, weighted by the amount of liquidity minted. Figure \ref{fig:ultimate_range_vs_rv} provides plots of these quantities, using marker shapes to distinguish token pairs and colors to distinguish fee tiers.

The figure suggests a positive correlation between the provision's range width and the (cleaned) RV. This is confirmed by the Spearman correlation test with permutation \cite{zwillinger1999crc}, whose null hypothesis is the absence of correlation between the variables. Furthermore, the result is stable if we normalize the tick range by the \texttt{tickSpacing}, suggesting it is not merely due to the mechanics of liquidity provision. The relationship is, however, weaker than what is typically observed in TradFi\footnote{This statement assumes that the TradFi relationship volatility-spread is a proxy of the DeFi relationship volatility-provision's range width.} -- see, for example, \cite{wyart2008relation}.

\section{Transition probabilities}
\label{sec:trans_prob}
This section focuses on the interdependence between events by studying the transition probabilities from one event type to another. Indeed, as our goal is to study the collective dynamics occurring on Uniswap, we wish to understand how (and if) a particular event plays a role in triggering another one. To this end, we study the transition probabilities. Specifically, given the events' time series $\{e_t\}_{t=1}^T$ measured in event-time, we estimate
\begin{equation}
\label{eqn:event_trans_prob}
    \mathbb{P}(e_{t+l}=E | e_t=E') \quad l=1,2,3,.. \quad E,E' \in \{\text{Mint, Burn, Swap zero, Swap ch}\}
\end{equation}

\begin{figure}[h]
	\centering
	\includegraphics[width=1\linewidth]{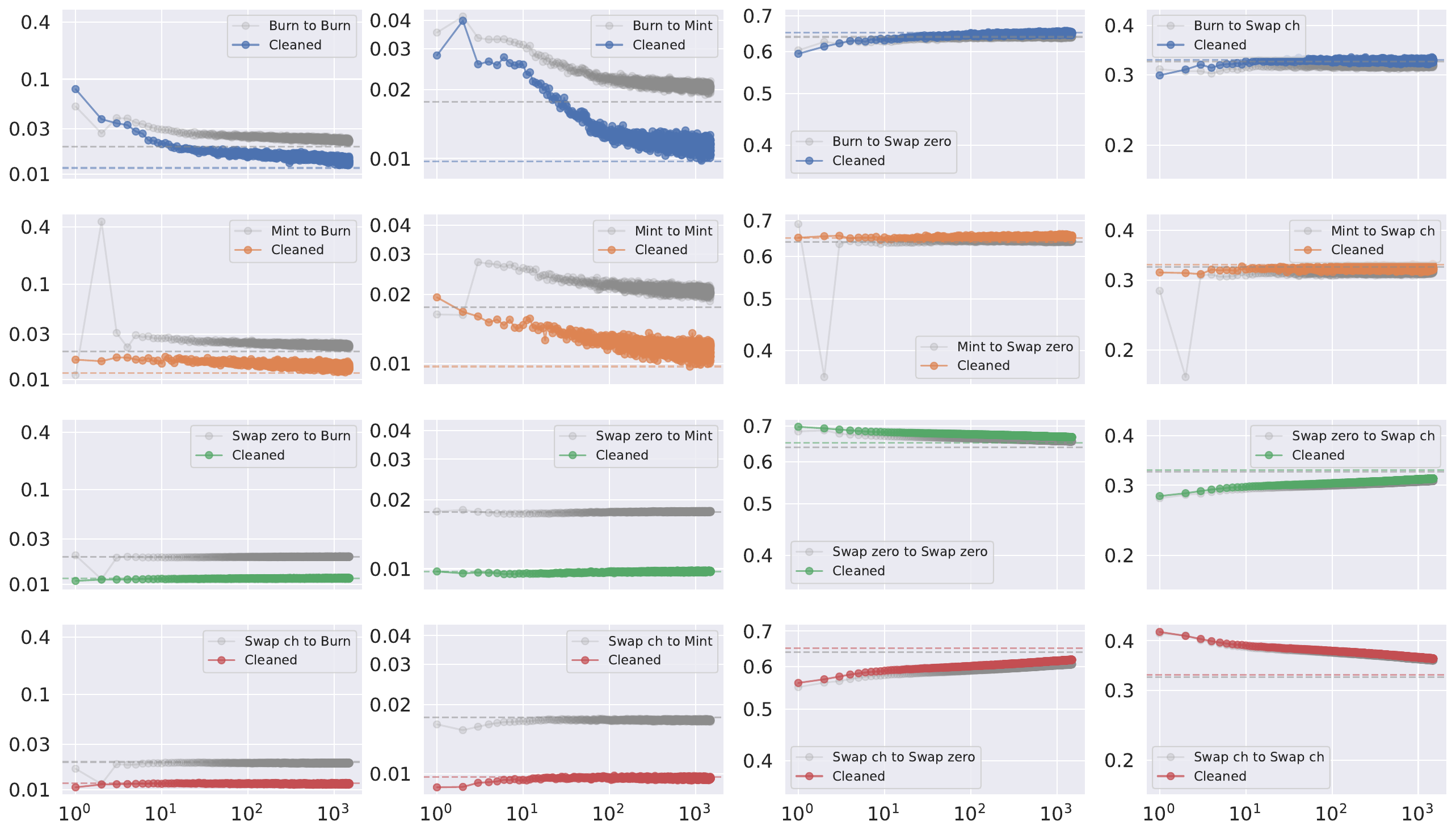}
	\caption{\texttt{USDC-WETH} 0.05\% pool - Empirical transition probabilities from one event type to another as a function of the lag $l$, see Eq. \eqref{eqn:event_trans_prob}. The gray lines are obtained from the whole dataset; the colored lines are obtained by removing MEV strategies from the dataset. By row, there is the conditioning event $e_t$; by column, there is the target $e_{t+l}$. The horizontal dashed lines indicate the unconditional probabilities $\mathbb{P}(e_{t+l}=E)$.}
	\label{img:stf_tr_probs_joined_usdc_weth_005_compressed}
\end{figure}

Swaps are partitioned into \textit{Swap zero} and \textit{Swap ch}, according to whether the corresponding tick movement is zero or not. Figure \ref{img:stf_tr_probs_joined_usdc_weth_005_compressed} shows the transition probabilities as a function of the lag in the \texttt{USDC-WETH} 0.05\% pool.

The gray lines are the conditional probabilities obtained from the whole event's time series; the dashed gray lines represent the unconditional probabilities. As expected, as the lag $l$ increases, all transition probabilities converge to their unconditional probabilities, although with differing rates. There appears to be a sharp peak at lag $2$ in the probabilities conditional to a mint event; similar peaks are in the probabilities of a burn conditional to a swap event. We conjecture that this is the result of the large volume of JIT events occurring in the pool. As already stated, a JIT event consists of a specific operational pattern occurring in the same block: mint, (often only one) swap, and burn. Furthermore, they account for about half of the liquidity provision events in almost all the pools. Therefore, it is reasonable that $\argmax_{E} \left( \mathbb{P}(e_{t+1}=E | e_t=Mint) - \mathbb{P}(e_{t+1}=E) \right)$ is Swap zero (the large amount of the active liquidity minted prevents from tick changes), $\argmax_{E} \left( \mathbb{P}(e_{t+2}=E | e_t=Mint) - \mathbb{P}(e_{t+2}=E) \right)$ is Burn, and so on. To have a more complete picture, we filter out the effect of MEV strategies by removing them from the data when computing the transition probabilities -- the output is plotted with the colored lines, contrasting with the gray lines that correspond to the conditional probabilities obtained without filtering MEV attacks. On a closer examination, all the events appear to elicit themselves with a probability greater than the unconditional one if they have just occurred -- that is, on the main diagonal of the plot, the continuous lines lie above the dashed ones. This pattern has already been observed in traditional finance and is known as the \textit{diagonal effect}\footnote{The definition of diagonal effect is  ``the tendency to see the same order types appear in succession''; see \cite{biais_hillon_spatt}}. Moreover, mint and burn orders excite each other. The strongest effect is from burn to mint, probably due to position adjustments. Conversely, swaps are not affected by the occurrences of liquidity provision events. Furthermore, swaps with and without tick changes seem to inhibit each other. One possible explanation for this is related to volatility clustering: during high-variance periods, swaps with tick changes are far more probable, whereas during low-volatility periods, swaps without tick movements frequently appear.
Finally, the above patterns are also observed in the other pools. Table \ref{sec:trans_prob} in Appendix \ref{app:oep} presents the results for each pool. Deviations from these stylized patterns appear just in five Stable and Synthetic pools, the same ones where volume long-memory is not observed. These are those with fewer events, making it likely that this ambiguity originates from the scarcity of data.

\section{Conclusion and Research Perspectives}
\label{sec:conclusion}
This study offers a first analysis, at the microstructural level, of Uniswap v3. This market has been chosen due to its academic and industrial relevance and the plethora of information and publicly available data. These features make it a perfect benchmark to analyze the impact of the DEXs' market structure on the statistical properties observed at high-frequency. Specifically, we focused on twenty-four highly traded pools spanning different types of underlying pairs and fee tiers. 
We analyzed the economic drivers behind the most widely used strategies, their impact on the pool activity, and the composition of the attackers. Specifically, we highlight that MEV strategies have a dramatic impact on the trading activity, and that the relevance of each specific strategy is strongly affected by the pool fee tier.

The core of this work focuses on price-related time series, where we detect significant patterns that partly follow those observed in TradFi. Nonetheless, the technical implementation of DEXs and the existence of MEV strategies lead to some differences with TradFi. Specifically, we find that swap returns are leptokurtic, with kurtosis (super)linearly decaying when a few swaps are aggregated. In DEXs, we observe a pattern known as echo swaps, characterized by two consecutive swaps placed by the same wallet in opposite directions -- a phenomenon not reported in TradFi. The autocorrelation of returns displays multiple peaks at lag $1$, reflecting the presence of echo swaps and reverse arbitrage, and at lags $2$ and $3$, corresponding to sandwich attacks. Contrastingly, in TradFi, only a single lag-one peak is typically observed. Order flow is correlated as well, but herding effects dominate over order splitting, and this pattern is persistent even when aggregating across pools. Liquidity often experiences sharp drops due to the Concentrated Liquidity mechanism; although not directly profitable, these drops trigger jumps in the marginal price, producing deviations from the CEX benchmark and artificially inflating volatility. Moreover, while in traditional markets volatility and spread are strongly related, in DEXs the correlation between volatility and the liquidity provision range is very weak. Finally, transition probabilities from one event to another also exhibit distinctive peaks that can be attributed to MEV activity. The appendix corroborates our findings with figures and tables describing the quantitative results.


From our study, there are many opportunities for future work. We cluster some of them into three main directions.
\begin{itemize}
	\item General questions about pools' agents, such as the presence of inter-pool patterns and correlation or causality effects between different pools. Furthermore, it would be extremely valuable to consider Ethereum Layers 2, like Arbitrium or Base. Due to their lower gas fees and the increased number of transactions validated per unit of time, they could reveal other intriguing statistical properties in the DeFi markets.
	\item Deeper analysis of the LTs' behavior and their activity. For example, a detailed analysis of router smart contracts, their strategies, and how they impact MEV searchers' activity, as well as a study of the impact of liquidity droughts.
	\item MEV searchers' advanced strategies and their impact -- e.g., an exhaustive explanation of the echo swaps economic drivers, as discussed in Subsection \ref{subsec:echo}. Moreover, what is the cause of long-memory ACF in the liquidity minted by JIT attacks? Finally, it may be interesting to carry out an analysis of the victims of MEV attacks. Are there agents who let them be jitted on purpose? Are MEV searchers, in turn, frequently victims of attacks by other searchers?
\end{itemize}

In conclusion, this study represents a step toward a comprehensive understanding of DEX microstructure, bridging the gap between traditional market analysis and DeFi. Nonetheless, additional effort is essential to fully understand the mechanisms driving DEX behavior. As this new ecosystem evolves, such foundational research will be critical for theoretical advancements and practical applications.

\section*{Acknowledgements}
    FL acknowledges support from the grant PRIN2022 DD N. 104 of February 2, 2022 ``Liquidity and systemic risks in centralized and decentralized markets'', codice proposta 20227TCX5W - CUP J53D23004130006 funded by the European Union NextGenerationEU through the Piano Nazionale di Ripresa e Resilienza (PNRR). SJ acknowledges support from the Natural Sciences and Engineering Research Council of Canada (RGPIN-2024-04317). We also thank Davide Barone for the insightful discussion about MEV.\\
\bibliographystyle{abbrv}
\bibliography{Bibliography}

\appendix
\section{Uniswap - Mathematical Details}
\label{app:uv3md}
This appendix is intended for the inexperienced reader approaching Uniswap markets for the first time. Here, we provide the mathematical rules behind Uniswap markets. First, we describe Uniswap v2. Then, we move to Uniswap v3, introducing Concentrated Liquidity and how it changes the game.
\subsection{Uniswap v2}
Uniswap v2 \cite{adams2020uniswap} is based on a specific type of Automated Market Makers named Constant Product Market Makers (CPMMs). In such systems, liquidity is stored in pools that operate independently, each containing a pair of tokens -let's denote them as X and Y. The latter is used as the numeraire. Denote the reserves in event-time $t$ as $R^X_t$ and $R^Y_t$. CPMMs allow three core events: swap, mint, and burn.\\

A swap refers to the exchange of the two tokens: the Liquidity Taker (LT) deposits $x_{swap}$ units of token X (or Y) into the pool and receives $y_{swap}$ units of token Y (or X) in return. The output $y_{swap}$ is computed to keep constant the product between reserves, net of the transaction fee paid by the LT. This is a fixed fraction $\phi\in(0,1)$ of the volume $x_{swap}$. It is intrinsic in the pool, so it cannot change (unless a decision of the pool governance is made in very special situations; however, we can neglect this chance and consider $\phi$ as a constant). Thus, the ultimate equation is:
\begin{equation}
\label{eq:0723_1839}
(R^X_t + (1-\phi)x_{swap})(R^Y_t - y_{swap}) = R^X_t R^Y_t = L_{tot, t}^2 \quad\implies\quad y_{swap} = x_{swap} \frac{(1-\phi)R^Y_t}{R^X_t + (1-\phi)x_{swap}}
\end{equation}
$L_{tot, t}$ is referred to as the total liquidity in the pool at time $t$. The transaction price is defined as the ratio $\frac{y_{swap}}{x_{swap}} = \frac{(1-\phi)R^Y_t}{R^X_t+(1-\phi)x_{swap}}$. If we neglect the fee and take the limit as $x\rightarrow0$, the transaction price approaches $\frac{R^Y_t}{R^X_t}$, which is referred to as the marginal price $S_t$, and it is used as the reference price. Moreover, it is worth noting that, in Uniswap v2, the fee paid by the LT is reinjected into the pool as additional liquidity. In this case, we talk about compounding fees. That is, even if in the price computation in Eq. \eqref{eq:0723_1839} the pair $(R^X_t + (1-\phi)x_{swap}, R^Y_t - y_{swap})$ is used, the actual pool reserves of token X are updated as $R^X_t + x_{swap}$ and $R^Y_t - y_{swap}$. Thus, the new marginal price is $\frac{R^X_t R^Y_t}{\left(R^X_t+(1-\phi)x_{swap} \right)(R^X_t+x_{swap})}$.\\

A mint operation involves supplying liquidity to the pool-i.e., adding a specified positive amount of both tokens X and Y, denoted as $x_{mint}$ and $y_{mint}$. Minting new liquidity requires the pool price to remain unchanged. Thus, for any liquidity addition, the following condition must hold:
\begin{equation}
\label{eq:0723_1740}
    \frac{R^Y_t}{R^X_t} = \frac{R^Y_t + y_{mint}}{ R^X_t + x_{mint} } \qquad \implies \qquad \frac{y_{mint}}{x_{mint}} = \frac{R^Y_t}{ R^X_t }
\end{equation}
An LP's stake in the pool is proportional to the liquidity they provide $L$. Elementary algebra then shows that the additional liquidity $L$ is given by $L=\sqrt{x_{mint} y_{mint}}$. Indeed, the constancy of price enforces $R^X_t y_{mint} = R^Y_t x_{mint}$, so we have that $R^X_t y_{mint} = R^Y_t x_{mint} = \sqrt{R^X_t y_{mint}} \sqrt{R^Y_t x_{mint}} = \sqrt{R^X_t R^Y_t} \sqrt{x_{mint} y_{mint}}$. Therefore:
\begin{equation}
\left(R^X_t + x_{mint}\right)\left(R^Y_t + y_{mint}\right) = \left(L_{tot, t}+L\right)^2 = \left(\sqrt{R^X_t R^Y_t}+L\right)^2 \qquad \implies \quad L = \sqrt{x_{mint} y_{mint}}
\end{equation}
For analyzing the position of an individual LP, it is often sufficient to relate their personal liquidity $L$ to the specific quantities of tokens deposited, disregarding the aggregate pool reserves. 

Furthermore, it is more convenient to describe the pool (respectively, the LP position) in terms of total liquidity and price $(L_{tot, t}, S_t)$ (respectively, the LP's stake and the price $(L, S_t)$) rather than token balances $(R^X_t, R^Y_t)$ (respectively, the tokens minted $(x_{mint}, y_{mint})$). Indeed, the former representation is more practical for analytical purposes and the smart contract reasons in terms of $L_{tot, t}$ and $S_t$. Temporal indices may be omitted for simplicity when no ambiguity arises. Finally, as for the burn operations, one can exploit the same computation as above, with $x_{burn}$ and $y_{burn}$ negative quantities.
\subsection{Uniswap v3}
\label{app:subsec_uv3}


The main limitation of Uniswap v2 is the lazy liquidity -i.e., the LPs are forced to provide liquidity and spread their tokens over the whole price interval $(0,+\infty)$. Uniswap v3 \cite{adams2021uniswap} is built upon and generalizes its previous version. The main innovation is the Concentrated Liquidity feature that allows LPs to select a specific position's range $R=[S_l, S_u)\subset\mathbb{R}_+$ within which to provide liquidity. Uniswap v2 can be viewed as the limiting case where $S_l=0$ and $S_u=\infty$. Thus, we first analyze the LP point of view. An LP is said to be active\footnote{In the literature, there is a duality in the definition of "active" LPs. Here, "active" means that the price is inside their price range $R$. In other contexts, such as the Just-in-Time liquidity analyzed in Subsection \ref{subsec:jit_liq}, "active" means that the LP acts as a reaction to a swap pending in the memory-pool. See Section \ref{subsec:jit_liq} of the main text for the full details. Nonetheless, we use the first meaning only in the current Appendix \ref{app:subsec_uv3}. Elsewhere, we always refer to the second meaning.} -contributing liquidity to the pool- whenever the current price falls within the chosen interval $R$. LP activity can be interpreted as follows: when the price lies within $R$, the LP holds both X and Y, facilitating trades for LTs. If the price drops below $S_l$ or rises above $S_u$, the LP’s entire liquidity is converted fully into X or Y, respectively, thus rendering them inactive. The total pool liquidity $L_{tot}$ and the total reserves become the sum of those of all active positions, and may fluctuate as swaps alter the price and thus the set of active LPs. Thus, the pool's liquidity is no longer constant. It can rather be viewed as a function of the price. This function is sometimes referred to as the liquidity profile. In the limiting Uniswap v2 case, all the provision positions have a range of $(0, \infty)$, thus the liquidity profile is a constant function.\\

To understand how the Concentrated Liquidity effectively increases the utility of the reserves, let's consider an LP minting $\tilde{x}$ units of X and $\tilde{y}$ units of Y into the position's range $R=[S_l, S_u)$. $\tilde{x}$ and $\tilde{y}$ are the real reserves -i.e., the tokens actually injected into the pool. However, the pool contract works in terms of artificial quantities known as the virtual reserves for X and Y, defined as $x_{mint} = \tilde{x} + \frac{L }{\sqrt{S_u}}$ and $y_{mint} = \tilde{y} + L\sqrt{S_l}$, where $L$ is the virtual liquidity minted, which is defined as the positive root of the following:
\begin{equation}
\label{eq:0723_1848}
x_{mint}y_{mint} = (\tilde{x} + \frac{L }{ \sqrt{S_u}}) (\tilde{y} + L\sqrt{S_l}) = L^2 \qquad \implies \qquad L = \frac{\tilde{x}\sqrt{S_l} + \frac{\tilde{y}}{\sqrt{S_u}} + \sqrt{\left(\tilde{x}\sqrt{S_l} - \frac{\tilde{y}}{\sqrt{S_u}}\right)^2 + 4\tilde{x}\tilde{y}}}{2\left(1-\sqrt{\frac{S_l}{S_u}}\right)}
\end{equation}
Specifically, $x_{mint}\ge\tilde{x}$, $y_{mint}\ge\tilde{y}$, and $L \ge \sqrt{\tilde{x}\tilde{y}}$ -that is, the smart contract assigns us more tokens than those actually provided. The equality is recovered in the limit cases $S_u=\infty$ and $S_l=0$ -i.e., when the concentrated liquidity feature is not exploited. When the liquidity is active, the virtual reserves ratio equals the marginal price -that is, $S = \frac{y_{mint}}{x_{mint}}$. Thus, we can rearrange $x_{mint}=\frac{L}{\sqrt{S}}$ and $y_{mint}=L\sqrt{S}$, which lead to $\tilde{x} = L\left( \frac{1}{\sqrt{S}} - \frac{1}{\sqrt{S_u}} \right)$ and $\tilde{y} =L\left(\sqrt{S} - \sqrt{S_l} \right)$. If $S<S_l$, then the real reserve of token Y is $\tilde{y}=0$. As for the price constraint, it will be active only if the liquidity touches the position's range -i.e., the price reaches $S_l$. Thus, $S_l = \frac{L\sqrt{S_l}}{x_{mint}}$, which leads to $\tilde{x} = L(\frac{1}{\sqrt{S_l}} - \frac{1}{\sqrt{S_u}})$. A similar reasoning holds if $S>S_u$. Summarizing, the mapping from $(L, S)$ to the true reserves $(\tilde{x}, \tilde{y})$ is described as:
\begin{equation}
\label{eq:0723_1859}
    \left(\tilde{x}, \tilde{y}\right) = \begin{cases} \left(L\left( \frac{1 }{ \sqrt{S_l}} - \frac{1 }{ \sqrt{S_u}} \right), 0\right) & if \ S<S_l \\ \left(L\left( \frac{1 }{ \sqrt{S}} - \frac{1 }{ \sqrt{S_u}} \right), L\left( \sqrt{S} - \sqrt{S_l} \right)\right) & if \ S_l\le S\le S_u \\  \left(0, L\left( \sqrt{S_u} - \sqrt{S_l} \right)\right) & if \ S>S_u \end{cases}
\end{equation}

The introduction of Concentrated Liquidity forced a technical adjustment of the pools' contracts that now implement a logic based on ticks. That is, the price space is partitioned into a discrete set of points defined as $1.0001^{j}$, with $j\in\mathbb{N}$. The integer exponents $j$ are known as ticks. So, at each price $S$ lying in the continuous set\footnote{Here, "continuous" has to be intended net of the numerical approximation.} $\mathbb{R}_+$, it is associated the tick $\lfloor \log_{1.0001} (S) \rfloor$ in the discrete set $\mathbb{N}$. When an LP provides liquidity, the range bounds $S_l$ and $S_u$ have to belong to the set $\{1.0001^{j\cdot ts} | j\in\mathbb{N}\}$, where $ts\in\mathbb{N}$ is the \texttt{tickSpacing} parameter (fixed by the pool's governance). Thus, the range bounds are identified with the corresponding ticks. Values in $\{j\cdot ts | j\in\mathbb{N}\}$ are called initializable ticks. Initializable ticks are said to be initialized if there is a liquidity provision position having that tick as the lower or upper bound.
Finally, it is worth underlining that the Concentrated Liquidity feature makes it infeasible to use the Uniswap v2 logic based on compounding fees. Thus, the fees collected by LPs in Uniswap v3 are stored separately and not reinjected into the pool. This way of collecting fees is referred to as non-compounding fees. This technicality could make the difference when looking at some mathematical properties of the markets. Thus, it is not certain that the results holding for Uniswap v2 pools are still valid in Uniswap v3.\\

Now, we take the point of view of an LT willing to place a swap order. Without loss of generality, assume that the LT wants to swap $x_{swap}$ units of token X with token Y, thus moving the price downward. Let $j$ be the price tick, $l$ the greatest initialized tick smaller than $j$, $u$ the smallest initialized tick greater than $j$, and $R^X$ and $R^Y$ the virtual reserves of the pool in $[l,u)$. By using the Uniswap v2 logic, we can obtain the output amount of a swap that does not pass the initialized ticks. Then, taking into account the non-compounding fees mechanism, it is easy to compute the price after such a swap. Ultimately, we recover the amount $x^*$ that, if traded, moves the marginal price to $l$:
\begin{equation}
    \text{New price v3: } \frac{R^X R^Y}{\left( R^X+(1-\phi)x_{swap} \right)^2} \qquad x^* = \sqrt{R^X} \frac{\sqrt{R^X + 1.0001^l \; R^Y} - \sqrt{R^X}}{1-\phi}
\end{equation}
As long as $x_{swap}\le x^*$, the price tick after the swap is still inside $[l, u)$. Thus, there is no liquidity adjustment and the output is computed as done in Eq. \eqref{eq:0723_1839}. Otherwise, the swap is split into two components. First, $x^*$ is traded. Then, the active liquidity is adjusted accordingly, causing an additional gas fee to be paid by the LT; lastly, the remaining part $x_{swap}-x^*$ is traded as if it were a new swap.\\

Finally, it is interesting to see how the non-compounding fee mechanism of Uniswap v3 causes some difference with respect to Uniswap v2. Indeed, let's consider a swap in a v3 pool. Assume that the swap does not pass any initialized tick -i.e., it does not cause a change in the liquidity profile. Then, the price impact, measured as the variation in the marginal price, is smaller, ceteris paribus, than that on the v2 pool. This is because more units of token X are put into the pool due to the compounding fees mechanism. For example, let us consider the order splitting problem, whose relevance is empirically analyzed in Subsection \ref{subsec:trade_sign} of the main text. Regarding the Uniswap v2 market, \cite{angeris2021analysis} shows that splitting an order is more expensive than executing it one-shot. That is, if an LT wants to swap $x_1+x_2$ units of token X for token Y, the output generated by trading them at once is, ceteris paribus, bigger than the sum of those obtained by first swapping $x_1$ and then $x_2$. Instead, when switching to Uniswap v3, this result does not hold. With straightforward computations, it is possible to show that the two quantities are precisely the same; thus, splitting the order is not more expensive than the one-shot execution\footnote{Nonetheless, when the gas fees are considered, we find that the split order is more expensive than the one-shot one.}.
\section{General Events Patterns}
\label{app:oep}
This appendix discusses some general patterns found in the event time series. First, we provide a detailed summary of the dataset used; then, we analyze the transition probabilities from one event type to another; finally, we have a look at the daily seasonality in the market activity. Figure \ref{tab:data_summary} summarizes the main properties of the pools considered in our analysis.
\begin{table}[h]
{\small
\centering
\begin{tabular}{lrrrrrrrr}
\toprule
\textbf{Pool} & \textbf{Start} & \textbf{\# Swaps} & \textbf{S/B} & \textbf{S/T} & \textbf{\# Mint} & \textbf{\# Burn} & \textbf{Wait B} & \textbf{\# Wallets} \\
\midrule
\texttt{USDC-WETH} 0.01\% & 2023-01-01 & 861,917 & 1.44 & 1.28 & 3,009 & 3,243 & 8.69 & 189,594 \\
\texttt{USDC-WETH} 0.05\% & 2023-01-01 & 4,115,698 & 1.56 & 2.97 & 75,449 & 83,461 & 1.95 & 824,769 \\
\texttt{USDC-WETH} 0.3\% & 2023-01-01 & 238,996 & 1.10 & 1.61 & 9,339 & 17,629 & 21.93 & 75,447 \\
\texttt{WETH-USDT} 0.01\% & 2023-02-22 & 1,917,101 & 1.61 & 1.43 & 6,970 & 7,441 & 4.05 & 404,850 \\
\texttt{WETH-USDT} 0.05\% & 2023-01-01 & 2,826,625 & 1.37 & 2.20 & 15,641 & 17,134 & 2.53 & 640,902 \\
\texttt{WETH-USDT} 0.3\% & 2023-01-01 & 420,329 & 1.10 & 2.46 & 13,757 & 16,524 & 13.00 & 149,109 \\
\texttt{WBTC-USDC} 0.05\% & 2023-01-01 & 205,205 & 1.14 & 1.09 & 2,812 & 3,058 & 28.61 & 36,743 \\
\texttt{WBTC-USDC} 0.3\% & 2023-01-01 & 91,080 & 1.07 & 1.09 & 5,236 & 6,404 & 56.02 & 14,035 \\
\texttt{WBTC-USDT} 0.05\% & 2023-03-31 & 66,454 & 1.12 & 1.08 & 586 & 603 & 76.41 & 17,784 \\
\texttt{WBTC-USDT} 0.3\% & 2023-01-01 & 96,804 & 1.06 & 1.13 & 3,068 & 3,529 & 54.75 & 17,791 \\
\midrule
\texttt{WBTC-WETH} 0.01\% & 2023-01-03 & 138,521 & 1.38 & 1.66 & 935 & 953 & 51.69 & 50,036 \\
\texttt{WBTC-WETH} 0.05\% & 2023-01-01 & 723,345 & 1.21 & 1.77 & 12,478 & 14,460 & 8.53 & 171,079 \\
\texttt{WBTC-WETH} 0.3\% & 2023-01-01 & 128,129 & 1.09 & 1.60 & 11,562 & 15,964 & 37.01 & 26,575 \\
\texttt{LINK-WETH} 0.3\% & 2023-01-01 & 241,087 & 1.07 & 1.25 & 15,211 & 18,416 & 21.18 & 35,812 \\
\texttt{MNT-WETH} 0.3\% & 2023-07-17 & 64,877 & 1.07 & 1.56 & 418 & 410 & 62.62 & 23,647 \\
\texttt{UNI-WETH} 0.3\% & 2023-01-01 & 156,546 & 1.05 & 1.17 & 6,401 & 7,990 & 32.71 & 22,188 \\
\midrule
\texttt{USDC-USDT} 0.01\% & 2023-01-01 & 685,117 & 1.12 & 19.94 & 9,095 & 10,032 & 8.42 & 232,696 \\
\texttt{USDC-USDT} 0.05\% & 2023-01-01 & 72,706 & 1.41 & 2.76 & 1,306 & 1,548 & 98.28 & 32,874 \\
\texttt{DAI-USDC} 0.01\% & 2023-01-01 & 162,182 & 1.06 & 131.43 & 917 & 924 & 33.79 & 55,482 \\
\texttt{DAI-USDT} 0.01\% & 2023-01-01 & 91,725 & 1.06 & 5.08 & 718 & 841 & 59.87 & 33,924 \\
\texttt{USDe-USDT} 0.01\% & 2023-11-23 & 144,103 & 1.16 & 4.58 & 2,778 & 2,854 & 22.94 & 53,749 \\
\midrule
\texttt{WETH-weETH} 0.01\% & 2024-01-08 & 45,166 & 1.16 & 18.20 & 207 & 231 & 65.90 & 20,457 \\
\texttt{wstETH-WETH} 0.01\% & 2023-01-24 & 153,790 & 1.10 & 11.74 & 3,545 & 3,823 & 35.01 & 36,203 \\
\texttt{WBTC-LBTC} 0.05\% & 2024-06-18 & 38,577 & 1.14 & 49.14 & 112 & 196 & 41.45 & 28,749 \\
\bottomrule
\end{tabular}
\caption{Summary of the dataset used. \textbf{Start} is for the first date in the pool (different from 2023-01-01 only if the pool has been initialized after this date. \textbf{S/B} and \textbf{S/T} are for the average number of swaps per block and tick change. \textbf{Wait B} is for the average number of blocks elapsed between two blocks containing an event related to the pool.}
\label{tab:data_summary}
}
\end{table}

\begin{table}[h]
{\small
\centering
\begin{adjustwidth}{-1.0cm}{}
\begin{tabular}{lrrrrrrrrrr}
\toprule
\textbf{Pool} & \textbf{Q0.1 S} & \textbf{Q0.5 S} & \textbf{Q0.9 S} & \textbf{std(r)} & \textbf{Kurt} & \textbf{RV} & \textbf{RV*} & \textbf{L min} & \textbf{Q0.5 S} & \textbf{L max} \\
\midrule
\texttt{USDC-WETH} 0.01\% & 1595.7 & 2545.6 & 3489.4 & 4.46 & 53.9 & 0.5594 & 0.4632 & -0.2 & 5.7 & 10.3 \\
\texttt{USDC-WETH} 0.05\% & 1599.2 & 2326 & 3544.6 & 0.06 & 2.2 & 0.3006 & 0.3006 & 5.8 & 7.3 & 9.6 \\
\texttt{USDC-WETH} 0.3\% & 1617.8 & 2432 & 3555.3 & 0.12 & 1.1 & 0.2208 & 0.2208 & 5.8 & 6.9 & 8.8 \\
\texttt{WETH-USDT} 0.01\% & 2302.1 & 2747.3 & 3623.4 & 2.05 & 69.0 & 0.5926 & 0.2802 & 0.9 & 6.0 & 10.1 \\
\texttt{WETH-USDT} 0.05\% & 1623.1 & 2247.8 & 3554.3 & 0.17 & 156.6 & 0.3037 & 0.3037 & 3.5 & 6.7 & 9.2 \\
\texttt{WETH-USDT} 0.3\% & 1654.6 & 2518.2 & 3614.5 & 0.09 & 0.9 & 0.2284 & 0.2284 & 6.0 & 6.7 & 8.6 \\
\texttt{WBTC-USDC} 0.05\% & 26266 & 54441 & 80225 & 5.19 & 19.5 & 0.2363 & 0.2363 & 5.7 & 10.0 & 13.6 \\
\texttt{WBTC-USDC} 0.3\% & 25685 & 57795 & 90956 & 0.18 & 0.2 & 0.1616 & 0.1616 & 9.5 & 10.9 & 13.0 \\
\texttt{WBTC-USDT} 0.05\% & 29299 & 72238 & 98477 & 3.43 & 0.8 & 0.3049 & 0.3016 & -inf & 10.3 & 13.3 \\
\texttt{WBTC-USDT} 0.3\% & 26784 & 61377 & 94456 & 0.21 & 1.6 & 0.1581 & 0.1581 & 9.2 & 10.8 & 12.6 \\
\midrule
\texttt{WBTC-WETH} 0.01\% & 20.852 & 23.396 & 25.065 & 3.22 & 18.8 & 0.1959 & 0.1322 & 22.0 & 27.2 & 30.7 \\
\texttt{WBTC-WETH} 0.05\% & 14.508 & 18.818 & 26.402 & 0.11 & 84.7 & 0.0723 & 0.0723 & 24.9 & 27.7 & 30.1 \\
\texttt{WBTC-WETH} 0.3\% & 14.331 & 17.982 & 25.349 & 0.08 & 2.2 & 0.0405 & 0.0405 & 26.6 & 28.0 & 29.3 \\
\texttt{LINK-WETH} 0.3\% & 141.2 & 213.53 & 264.29 & 0.15 & 2.4 & 0.2071 & 0.2071 & 22.6 & 23.9 & 26.2 \\
\texttt{MNT-WETH} 0.3\% & 2770.2 & 3780.1 & 4374.2 & 0.26 & 0.7 & 0.2754 & 0.2754 & 23.4 & 23.8 & 25.4 \\
\texttt{UNI-WETH} 0.3\% & 239.62 & 341.17 & 409.04 & 0.19 & 0.8 & 0.3116 & 0.3116 & 22.3 & 23.6 & 26.1 \\
\midrule
\texttt{USDC-USDT} 0.01\% & 0.99908 & 0.99988 & 1.0007 & 1.89 & 135.0 & 0.0147 & 0.0102 & 8.5 & 16.6 & 18.1 \\
\texttt{USDC-USDT} 0.05\% & 0.91193 & 0.99859 & 1.0004 & 5.23 & 9.5 & 0.0030 & 0.0030 & 9.2 & 15.0 & 16.9 \\
\texttt{DAI-USDC} 0.01\% & 0.99996 & 1 & 1.0001 & 0.28 & 80.0 & 0.0002 & 0.0000 & 2.0 & 11.7 & 12.1 \\
\texttt{DAI-USDT} 0.01\% & 0.99924 & 0.99991 & 1.0007 & 18.05 & 11.0 & 0.6527 & 0.0080 & -inf & 9.4 & 11.5 \\
\texttt{USDe-USDT} 0.01\% & 0.99885 & 1.0012 & 1.0019 & 1.12 & 41.9 & 0.0022 & 0.0022 & 2.0 & 9.2 & 11.3 \\
\midrule
\texttt{WETH-weETH} 0.01\% & 1.043 & 1.0469 & 1.0544 & 3.34 & 6.0 & 0.0287 & 0.0285 & -inf & 24.7 & 25.6 \\
\texttt{wstETH-WETH} 0.01\% & 1.1389 & 1.1669 & 1.1859 & 4.10 & 24.1 & 0.0027 & 0.0005 & -inf & 25.2 & 26.2 \\
\texttt{WBTC-LBTC} 0.05\% & 0.99666 & 0.99834 & 0.99966 & 0.91 & 5.4 & 0.3671 & 0.0002 & 4.2 & 12.7 & 13.3 \\
\bottomrule
\end{tabular}
\caption{Summary of the price and liquidity time series. \textbf{Q0.1}, \textbf{Q0.5}, and \textbf{Q0.9} are for the 0.1, 0.5, and 0.9 quantiles. \textbf{S} and \textbf{L} are for the (marginal) price and $\log_{10}$(Liquidity) time series. \textbf{std(r)} and \textbf{Kurt} are for the standard deviation ($\cdot10^2$) and the excess kurtosis ($\cdot10^{-3}$) of the returns unconditional distribution. \textbf{RV} is for the Realized Variance (annualized) computed using the marginal prices sampled from the entire period every 10 minutes. \textbf{RV*} is the same, but computed by only considering prices such that the corresponding active liquidity is at least equal to the 5\% quantile of the historical active liquidity for that pool. Extremely small values of the minimum $\log_{10}$(Liquidity) in the \textbf{L min} column highlight the presence of the liquidity droughts discussed in the main text; their effect is highlighted by the difference between the columns \textbf{RV} and \textbf{RV*}.}
\label{tab:SrL_summary}
\end{adjustwidth}
}
\end{table}

Section \ref{sec:trans_prob} of the main text points out that the transition probabilities are strongly affected by the presence of MEV transactions, especially JIT strategies. Once cleaned for these, we notice mainly four patterns: diagonal effect, mint is more likely to occur after a burn, swap occurrences are independent of liquidity provision events, and swaps with or without a tick change inhibit each other. Table \ref{tab:trans_prob} summarizes occurrences or deviations from these patterns.
\begin{table}[h]
{\small
\centering
\begin{adjustwidth}{-0.5cm}{}
\begin{tabular}{lrrrr|lrrrr}
\toprule
\textbf{Pool}    & \multicolumn{1}{c}{\textbf{\begin{tabular}[c]{@{}c@{}}Diag\\ Eff.\end{tabular}}} & \multicolumn{1}{c}{\textbf{\begin{tabular}[c]{@{}c@{}}Burn\\ Mint\end{tabular}}} & \multicolumn{1}{c}{\textbf{\begin{tabular}[c]{@{}c@{}}Indep.\\ Swaps\end{tabular}}} & \multicolumn{1}{c|}{\textbf{\begin{tabular}[c]{@{}c@{}}Swaps\\ Block\end{tabular}}} & \textbf{Pool} & \multicolumn{1}{c}{\textbf{\begin{tabular}[c]{@{}c@{}}Diag\\ Eff.\end{tabular}}} & \multicolumn{1}{c}{\textbf{\begin{tabular}[c]{@{}c@{}}Burn\\ Mint\end{tabular}}} & \multicolumn{1}{c}{\textbf{\begin{tabular}[c]{@{}c@{}}Indep.\\ Swaps\end{tabular}}} & \multicolumn{1}{c}{\textbf{\begin{tabular}[c]{@{}c@{}}Swaps\\ Block\end{tabular}}} \\
\midrule
\texttt{USDC-WETH} 0.01\% & Yes & Yes & Yes & Yes & \texttt{WBTC-WETH} 0.3\%    & Yes & Yes & Yes & Yes \\
\texttt{USDC-WETH} 0.05\% & Yes & Yes & Yes & Yes & \texttt{LINK-WETH} 0.3\%    & Yes & Yes & Yes & Yes \\
\texttt{USDC-WETH} 0.3\%  & Yes & Yes & Yes & Yes & \texttt{MNT-WETH} 0.3\% & No & No  & Yes & Yes \\
\texttt{WETH-USDT} 0.01\% & Yes & Yes & Yes & Yes & \texttt{UNI-WETH} 0.3\% & Yes & Yes & Yes & Yes \\ \cline{6-10} 
\texttt{WETH-USDT} 0.05\% & Yes & Yes & Yes & Yes & \texttt{USDC-USDT} 0.01\%   & Yes & Yes & Yes & Yes \\
\texttt{WETH-USDT} 0.3\%  & Yes & Yes & Yes & Yes & \texttt{USDC-USDT} 0.05\%   & Yes & Yes & Yes & Yes \\
\texttt{WBTC-USDC} 0.05\% & Yes & Yes & Yes & Yes & \texttt{DAI-USDC} 0.01\%    & No & No  & Yes & No \\
\texttt{WBTC-USDC} 0.3\%  & Yes & Yes & Yes & Yes & \texttt{DAI-USDT} 0.01\%    & No & Yes & Yes & Yes \\
\texttt{WBTC-USDT} 0.05\% & Yes & Yes & Yes & Yes & \texttt{USDe-USDT} 0.01\%   & Yes & Yes & Yes & Yes \\ \cline{6-10} 
\texttt{WBTC-USDT} 0.3\%  & Yes & Yes & Yes & Yes & \texttt{WETH-weETH} 0.01\%  & No & No  & Yes & No \\ \cline{1-5}
\texttt{WBTC-WETH} 0.01\% & Yes & Yes & Yes & Yes & \texttt{wstETH-WETH} 0.01\% & Yes & Yes & Yes & Yes \\
\texttt{WBTC-WETH} 0.05\% & Yes & Yes & Yes & Yes & \texttt{WBTC-LBTC} 0.05\%   & No & Yes & No  & No \\
\bottomrule
\end{tabular}
\caption{Transition Probability Patterns - The table shows which pools clearly exhibit a given pattern. \textbf{Diag Eff.} is for the diagonal effect; \textbf{Burn Mint} is for the mint events elicitation by burn; \textbf{Indep. Swaps} represents the swaps' independence of liquidity provision events; \textbf{Swaps Block} represents swaps with or without a tick change that inhibit each other.}
\label{tab:trans_prob}
\end{adjustwidth}
}
\end{table}
\subsection{Intraday Patterns}
\label{app:intraday}
Now, we focus on the intraday patterns. The analysis concerns the market activity measured as the number of events or price volatility. DEXs are continuously open 24 hours\footnote{According to this perspective, Foreign Exchanges (FXs) are the TradFi market most similar to DEXs. As intra-day patterns corresponding to different market sessions are noticed in FXs, we are motivated to investigate similar behavior in our dataset.} and trading is possible also during holidays.  Thus, it makes sense to search for intraday patterns driven by changes in worldwide trading activity during the day. By examining these patterns, we can uncover the underlying mechanisms of liquidity, volatility, and investor behavior unique to DeFi platforms.

To study the intraday pattern, we first divide a day into time windows of 10 minutes and compute the percentage quantities of the number of events for each window. These can be viewed as discrete observations of an underlying market intensity process. Then, we project onto the Fourier space and perform an Functional Principal Components Analysis (FPCA) \cite{ramsay2013functional} to obtain the principal modes of variation in the data.

\begin{figure}[h]
	\centering
	\includegraphics[width=1\linewidth]{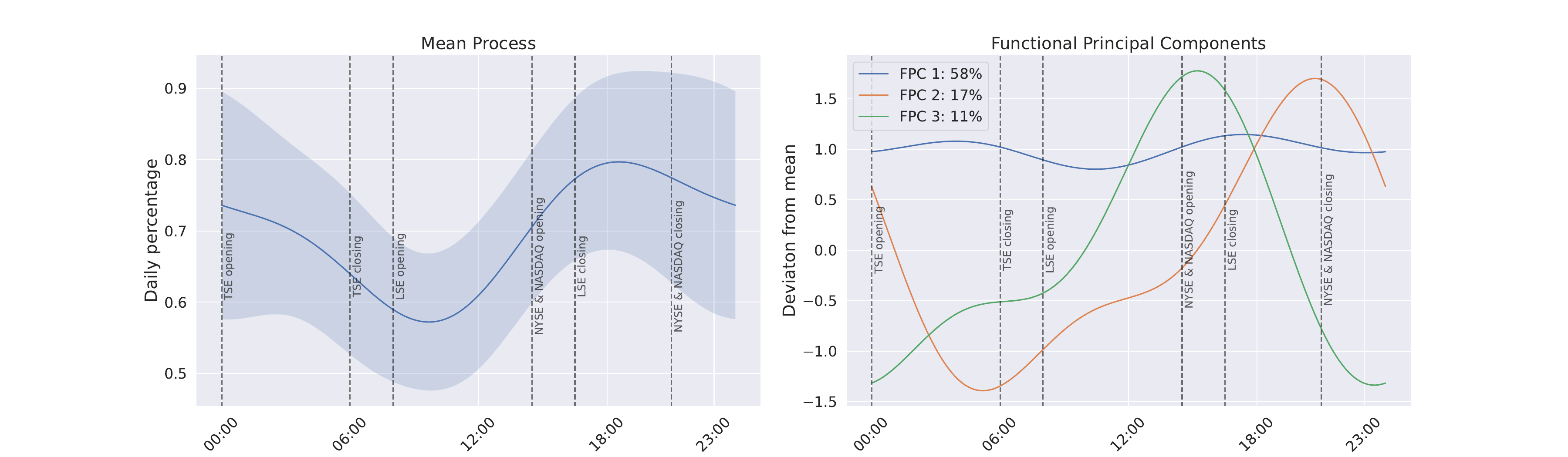}
	\caption{The left plot shows the mean process from the projection of the daily percentage market activity onto the first five Fourier bases. The shaded area corresponds to $\pm$ one standard deviation. On the right plot, there are the first 3 functional principal components with the corresponding explained variance ratio in the legend. They explain in total about 86\% of the variation present in the data. The horizontal axis represents normalized daily time (time zone UTC). The vertical dashed lines correspond to the opening and closing of the main market sessions. TSE is for the Tokyo Stock Exchange; LSE is for the London Stock Exchange; NYSE is for the New York Stock Exchange.}
	\label{img:mean_FPCA}
\end{figure}

Figure \ref{img:mean_FPCA} shows, as an example, the result for the  \texttt{USDC-WETH} 0.05\% pool, while the results of the other pools are in Figure \ref{img:fpca_mean_thr200k} of the Appendix \ref{app:oep}.  The left panel shows the mean of the market activity process and the right panel shows the first three Functional Principal Components (FPCs) that explain 86\% of the total variance.

As for the FPCs, the first component explains 58\% of the variance and is roughly flat. Thus, it corresponds to the variation of the market activity process between different days. Instead, the other FPCs can indicate the impact on the market intensity process of the European and American sessions (second component) and the Tokyo and European sessions (third component).

As for the mean process, we find that the \texttt{USDC-WETH} 0.05\% pool follows a common pattern shared by several pools, mostly in Normal and Volatile clusters. They are characterized by a decrease in market activity between 8 and 10 am UTC time (corresponding to the close of Sydney and Tokyo sessions) and an increase between 3 and 5 pm. As different pools share this pattern, we can conclude it does not appear to be a matter of chance. 

Moreover, a similar result holds when changing the window length or considering the returns' volatility as a proxy for the market activity.
To focus this point, we extend the previous analysis by showing the results obtained for all the examined pools and introducing the Modulated Realized Variance (MRV) as a proxy of the market activity. Specifically, Figure \ref{img:fpca_mean_thr200k} shows the mean value (across different days) of the projection of the percentage market activity in the Fourier space. It has been normalized by dividing by the daily mean. In this way, the activity curves of all the pools have the same mean and are comparable. As shown, most of the Normal pools share a common pattern made up of a drop in the pool activity around 08 am (UTC time) and an increase between 03 and 05 pm. The same behavior is noticeable in some of the Volatile and Stable pools, even if it is much lighter.

\begin{figure}[h!]
\centering
    \includegraphics[scale=0.30]{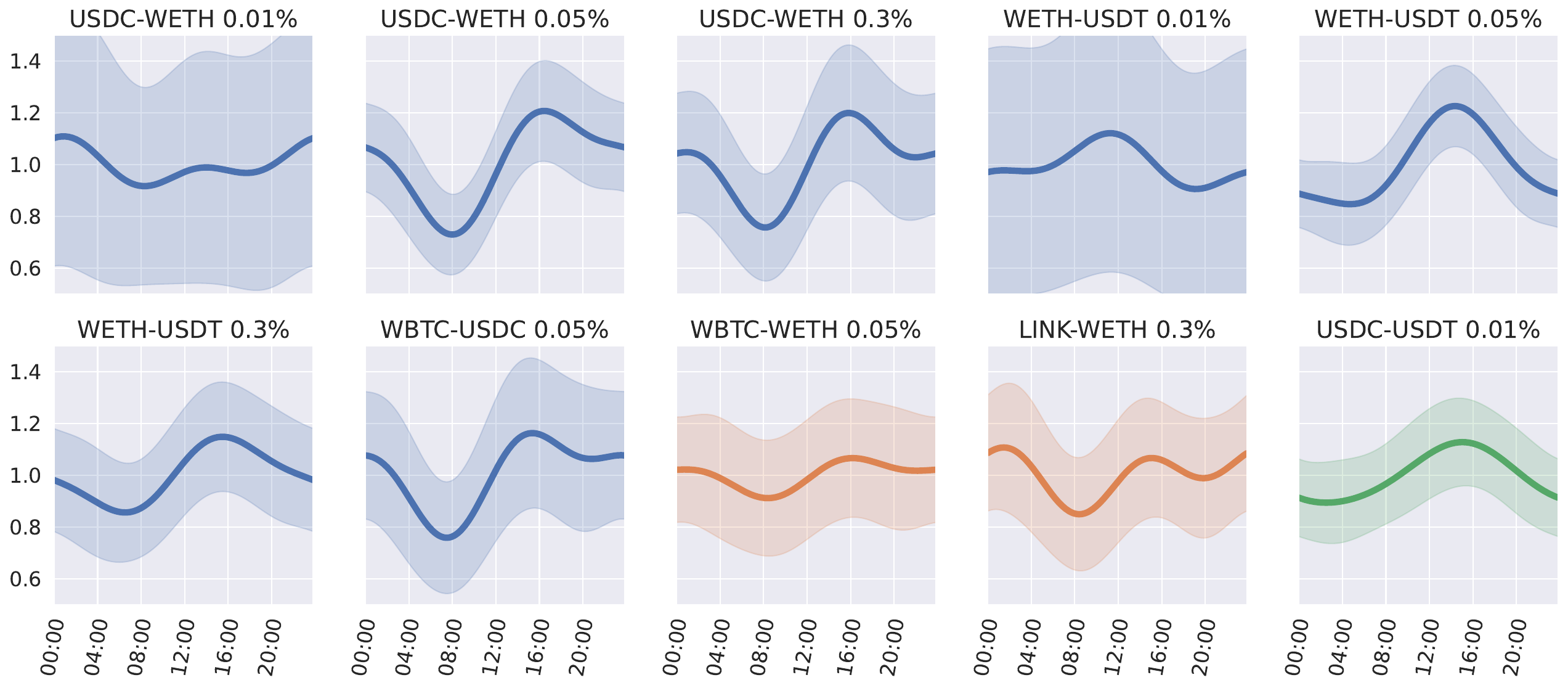}
    \vspace{-.3cm}
	\caption{Market activity - comparison between pools (with 200,000 events, at least). The market activity is measured as the mean rescaled percentages of the number of events projected onto the Fourier space. Normal, Volatile, and Stable pools are highlighted in blue, orange, and green colours. The shadow area corresponds to $\pm \frac{1}{2}$ standard deviation.}
\label{img:fpca_mean_thr200k}
\end{figure}

\begin{figure}[h!]
\centering
    \includegraphics[scale=0.30]{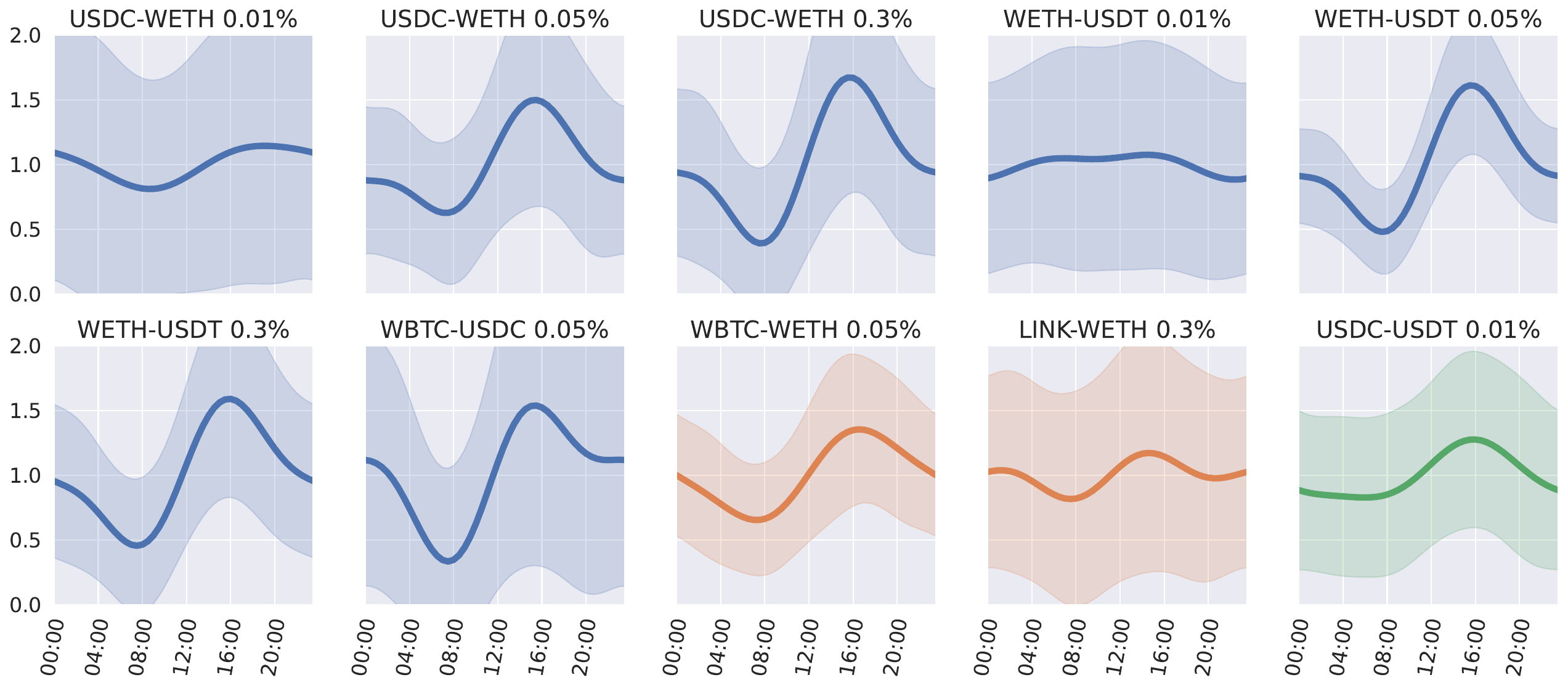}
    \vspace{-.3cm}
	\caption{Daily market activity measured in terms of the MRV. Normal, Volatile, and Stable pools are highlighted in blue, orange, and green. The shadow area is for the mean $\pm \frac{1}{2}$ standard deviation.}
\label{img:fpca_mean_thr200k_mrv}
\end{figure}

Another analysis about intraday patterns has been carried out by using the MRV as a proxy of the market activity. Specifically, the MRV is used to filter the integrated variance when the data are affected by noise -see \cite{podolskij2009estimation} for further details. The remainder of the pipeline is the same as before. The mean process is shown in Figure \ref{img:fpca_mean_thr200k_mrv} and is roughly coherent with what is observed in Figure \ref{img:fpca_mean_thr200k}.
\\

\clearpage
\section{Swap Events - Quantitative Results}
\label{app:seqr}
This section provides the quantitative results for the swap-related time series. Specifically, we first analyze the kurtosis of the log returns distribution. Next, we discuss the first lag in the returns ACF. Then, we move to the long-memory property exhibited by some time series based on swap events: absolute return, volume, and sign. Finally, we discuss the relationship between volume and variance.

Section \ref{subsec:fat_t} of the main text discusses the kurtosis and its scaling properties as the swaps aggregation size increases. When aggregating more swaps, we recover the power-law behavior typically observed in TradFi. Table \ref{tab:kurt} shows the quantitative results, pool-by-pool, of the power-law fit and the mesokurtosis test.
\begin{table}[h]
{\footnotesize
\centering
\begin{adjustwidth}{-1cm}{}
\begin{tabular}{lrrrrr|lrrrrr}
\toprule
\textbf{Pool}    & \textbf{log(A)} & \textbf{p}  & \textbf{R2} & \textbf{\begin{tabular}[c]{@{}r@{}}p-val\\ \textgreater 0.01\end{tabular}} & \multicolumn{1}{l|}{\textbf{\begin{tabular}[c]{@{}l@{}}p-val\\ \textgreater 0.05\end{tabular}}} & \textbf{Pool} & \textbf{log(A)} & \textbf{p} & \textbf{R2}   & \textbf{\begin{tabular}[c]{@{}r@{}}p-val\\ \textgreater 0.01\end{tabular}} & \multicolumn{1}{l}{\textbf{\begin{tabular}[c]{@{}l@{}}p-val\\ \textgreater 0.05\end{tabular}}} \\
\midrule
\texttt{USDC-WETH} 0.01\% & \underline{ 9.65} & \underline{ 0.77} & \underline{ 0.76}  & \underline{ 10.1} & \underline{ 16.2} & \texttt{WBTC-WETH} 0.3\%    & 3.26 & 0.26 & 0.66 & 1.80 & 2.83 \\
\texttt{USDC-WETH} 0.05\% & \underline{ 5.31} & \underline{ 0.38} & \underline{ 0.84}  & \underline{ 2.14} & \underline{ 2.70} & \texttt{LINK-WETH} 0.3\%    & 2.80 & 0.20 & 0.68 & 1.56 & 1.68 \\
\texttt{USDC-WETH} 0.3\%  & 2.83 & 0.20 & 0.67 & 1.13 & 1.13 & \texttt{MNT-WETH} 0.3\% & 3.74 & 0.32 & 0.71 & 2.81 & 3.55 \\
\texttt{WETH-USDT} 0.01\% & \underline{ 10.80} & \underline{ 0.81} & \underline{ 0.89}  & \underline{ 21.47} & \underline{ 25.55} & \texttt{UNI-WETH} 0.3\% & 3.93 & 0.32 & 0.64 & 3.03 & 3.17 \\ \cline{7-12} 
\texttt{WETH-USDT} 0.05\% & 7.00 & 0.57 & 0.70 & 3.08 & 4.21 & \texttt{USDC-USDT} 0.01\%   & \textbf{11.12}  & \textbf{0.85} & \textbf{0.93} & \textbf{16.12} & \textbf{20.80} \\
\texttt{WETH-USDT} 0.3\%  & \underline{ 3.62} & \underline{ 0.27} & \underline{ 0.81}  & \underline{ 1.86} & \underline{ 3.80} & \texttt{USDC-USDT} 0.05\%   & 7.80 & 0.83 & 0.61 & 10.98 & 17.08 \\
\texttt{WBTC-USDC} 0.05\% & 8.76 & 0.95 & 0.70 & 2.10 & 2.90 & \texttt{DAI-USDC} 0.01\%    & \textbf{12.16}  & \textbf{1.21} & \textbf{0.90} & \textbf{7.10} & \textbf{8.05} \\
\texttt{WBTC-USDC} 0.3\%  & 2.06 & 0.11 & 0.40 & 1.19 & 1.40 & \texttt{DAI-USDT} 0.01\%    & \underline{ 9.58} & \underline{ 0.95}    & \underline{ 0.80}    & \underline{ 8.18} & \underline{ 15.32} \\
\texttt{WBTC-USDT} 0.05\% & 7.43 & 0.67 & 0.68 & 26.02 & 28.21 & \texttt{USDe-USDT} 0.01\%   & \underline{ 9.77} & \underline{ 1.01}    & \underline{ 0.79}    & \underline{ 4.93} & \underline{ 4.93} \\ \cline{7-12} 
\texttt{WBTC-USDT} 0.3\%  & 2.44 & 0.18 & 0.74 & 1.76 & 2.07 & \texttt{WETH-weETH} 0.01\%  & \textbf{9.39}   & \textbf{0.83} & \textbf{0.92} & \textbf{0.01} & \textbf{0.01} \\ \cline{1-6}
\texttt{WBTC-WETH} 0.01\% & 7.29 & 0.54 & 0.64 & 0.01 & 0.01 & \texttt{wstETH-WETH} 0.01\% & \underline{ 9.28} & \underline{ 0.72}    & \underline{ 0.86}    & \underline{ 0.00} & \underline{ 0.00} \\
\texttt{WBTC-WETH} 0.05\% & 4.28 & 0.32 & 0.60 & 2.19 & 3.52 & \texttt{WBTC-LBTC} 0.05\%   & \textbf{9.84}   & \textbf{0.90} & \textbf{0.96} & \textbf{0.00} & \textbf{0.00} \\
\bottomrule
\end{tabular}
\caption{Kurtosis as a function of the number of aggregated swaps. $\log(A)$ and $p$ are the coefficients fitted from the regression $K(n) \sim A \cdot n^{-p}$. \textbf{R2} is for the R2 score of the linear regression $\log(K(n)) \sim \log(A) - p\log(n)$. \textbf{p-val > 0.01} and \textbf{p-val > 0.05} equal the maximum aggregation size $n$ such that the test in \cite{anscombe1983distribution} has p-value greater than 0.01 or 0.05, normalized by the mean number of swaps in a day. Pools where the R2 score of the power-law regression is greater than 0.9 (0.75) are in bold (underlined).}
\label{tab:kurt}
\end{adjustwidth}
}
\end{table}

As outlined in Section \ref{subsec:ret_acf} of the main text, the returns ACF of several pools shows a huge negative peak at the first lag. This is robust to sandwich attacks. Our guess is that it is related to the action of reverse arbitrageurs. To test for this, we disentangle the first lag ACF by conditioning on $abs(r_{t-1}) > \phi$ and $abs(r_{t-1}) \le \phi$, where $r_{t-1}$ is the return induced by the transaction in swap-time $t-1$, $abs(\cdot)$ is the absolute value, and $\phi$ is the fee tier. The results of the decomposition are in Table \ref{tab:l1d}.
\begin{table}[h]
{\small
\centering
\begin{tabular}{lrrr|lrrr}
\toprule
\textbf{Pool}    & \textbf{\begin{tabular}[c]{@{}r@{}}ACF\\ (1)\end{tabular}} & \textbf{\begin{tabular}[c]{@{}r@{}}ACF\\ (1 | g)\end{tabular}} & \textbf{\begin{tabular}[c]{@{}r@{}}ACF\\ (1 | l)\end{tabular}} & \textbf{Pool} & \textbf{\begin{tabular}[c]{@{}r@{}}ACF\\ (1)\end{tabular}} & \textbf{\begin{tabular}[c]{@{}r@{}}ACF\\ (1 | g)\end{tabular}} & \textbf{\begin{tabular}[c]{@{}r@{}}ACF\\ (1 | l)\end{tabular}} \\
\midrule
\texttt{USDC-WETH} 0.01\% & -0.014 & -0.015 & \textbf{0.000} & \texttt{WBTC-WETH} 0.3\% & 0.056 & \textbf{-0.142} & \textbf{0.256} \\
\texttt{USDC-WETH} 0.05\% & -0.042 & \textbf{-0.071} & \textbf{0.038} & \texttt{LINK-WETH} 0.3\% & -0.158 & \textbf{-0.414} & \textbf{0.211} \\
\texttt{USDC-WETH} 0.3\% & 0.029 & \textbf{-0.210} & \textbf{0.296} & \texttt{MNT-WETH} 0.3\% & -0.000 & \textbf{-0.122} & \textbf{0.166} \\
\texttt{WETH-USDT} 0.01\% & -0.026 & \textbf{-0.032} & \textbf{-0.002} & \texttt{UNI-WETH} 0.3\% & -0.077 & \textbf{-0.311} & \textbf{0.317} \\ \cline{5-8} 
\texttt{WETH-USDT} 0.05\% & -0.035 & \textbf{-0.046} & \textbf{0.010} & \texttt{USDC-USDT} 0.01\% & -0.096 & \textbf{-0.123} & \textbf{-0.001} \\
\texttt{WETH-USDT} 0.3\% & 0.016 & \textbf{-0.194} & \textbf{0.259} & \texttt{USDC-USDT} 0.05\% & -0.494 & \textbf{-0.536} & \textbf{0.007} \\
\texttt{WBTC-USDC} 0.05\% & -0.133 & \textbf{-0.139} & \textbf{0.003} & \texttt{DAI-USDC} 0.01\% & -0.497 & -0.497 & \textbf{-0.021} \\
\texttt{WBTC-USDC} 0.3\% & 0.003 & \textbf{-0.231} & \textbf{0.385} & \texttt{DAI-USDT} 0.01\% & -0.003 & -0.004 & -0.005 \\
\texttt{WBTC-USDT} 0.05\% & -0.218 & \underline{-0.226} & \textbf{0.022} & \texttt{USDe-USDT} 0.01\% & -0.139 & \textbf{-0.174} & \textbf{-0.008} \\ \cline{5-8} 
\texttt{WBTC-USDT} 0.3\% & 0.020 & \textbf{-0.128} & \textbf{0.301} & \texttt{WETH-weETH} 0.01\% & -0.496 & \textbf{-0.682} & \textbf{0.027} \\ \cline{1-4}
\texttt{WBTC-WETH} 0.01\% & -0.270 & -0.271 & \textbf{0.004} & \texttt{wstETH-WETH} 0.01\% & -0.310 & \textbf{-0.359} & \textbf{0.008} \\
\texttt{WBTC-WETH} 0.05\% & -0.089 & \textbf{-0.124} & \textbf{0.074} & \texttt{WBTC-LBTC} 0.05\% & -0.077 & -0.085 & \textbf{0.011} \\
\bottomrule
\end{tabular}
\caption{ACF lag 1 decomposition - The ACF at lag 1 \textbf{ACF(1)} is decomposed by conditioning on the previous return's magnitude bigger or smaller than the pool fee tier, respectively, \textbf{ACF(1 | g)} and \textbf{ACF(1 | l)}. The conditional ACFs outside the 99\% (95\%) confidence interval of the unconditional ACF are in bold (underlined).}
\label{tab:l1d}
}
\end{table}

Next, we analyze the quantitative results for long-memory. Specifically, Table \ref{tab:lm_rrts} examines long-memory in time series based on swap events by showing the power-law fit to the ACF of absolute return, volume, and sign.
\begin{table}[h]
\centering
\begin{tabular}{l|lll|lll|lll}
\toprule
\multirow{2}{*}{\textbf{Pool}} & \multicolumn{3}{c|}{\textbf{Abs}}  & \multicolumn{3}{c|}{\textbf{Vol}}  & \multicolumn{3}{c}{\textbf{Sign}} \\
 & \textbf{log(A)} & \textbf{p} & \textbf{R2}   & \textbf{log(A)} & \textbf{p} & \textbf{R2}   & \textbf{log(A)} & \textbf{p} & \textbf{R2}   \\
\midrule
\texttt{USDC-WETH} 0.01\%   & -6.38 & 0.09 & 0.03 & \textbf{-3.18} & \textbf{0.05} & \textbf{0.22} & -2.86 & 0.59 & 0.27 \\
\texttt{USDC-WETH} 0.05\%   & \textbf{-1.42} & \textbf{0.20} & \textbf{0.88} & \textbf{-1.78} & \textbf{0.32} & \textbf{0.95} & \textbf{-3.35} & \textbf{0.22} & \textbf{0.91} \\
\texttt{USDC-WETH} 0.3\%    & \textbf{0.16}  & \textbf{0.51} & \textbf{0.94} & \textbf{0.36}  & \textbf{0.51} & \textbf{0.82} & -5.32 & 0.06 & 0.00 \\
\texttt{WETH-USDT} 0.01\%   & -5.12 & 0.12 & 0.15 & \textbf{-3.64} & \textbf{0.11} & \textbf{0.52} & -2.31 & 0.68 & 0.43 \\
\texttt{WETH-USDT} 0.05\%   & \textbf{-2.85} & \textbf{0.21} & \textbf{0.90} & \textbf{-2.15} & \textbf{0.17} & \textbf{0.94} & \textbf{-3.62} & \textbf{0.26} & \textbf{0.78} \\
\texttt{WETH-USDT} 0.3\%    & \textbf{-0.07} & \textbf{0.41} & \textbf{0.92} & \textbf{-1.50} & \textbf{0.21} & \textbf{0.95} & -3.93 & 0.09 & 0.06 \\
\texttt{WBTC-USDC} 0.05\%   & -5.12 & 0.35 & 0.11 & \textbf{-1.88} & \textbf{0.17} & \textbf{0.79} & -5.25 & 0.19 & 0.03 \\
\texttt{WBTC-USDC} 0.3\%    & \textbf{-0.19} & \textbf{0.33} & \textbf{0.91} & \textbf{-1.34} & \textbf{0.19} & \textbf{0.87} & -5.77 & -0.03 & 0.00 \\
\texttt{WBTC-USDT} 0.05\%   & -1.92 & 0.34 & 0.31 & \textbf{-1.17} & \textbf{0.16} & \textbf{0.75} & -3.66 & 0.30 & 0.09 \\
\texttt{WBTC-USDT} 0.3\%    & \textbf{-0.14} & \textbf{0.46} & \textbf{0.90} & \textbf{-1.39} & \textbf{0.22} & \textbf{0.86} & -4.37 & 0.18 & 0.02 \\

\midrule

\texttt{WBTC-WETH} 0.01\%   & -1.90 & 0.47 & 0.39 & \textbf{-1.41} & \textbf{0.39} & \textbf{0.74} & \textbf{-1.30} & \textbf{0.25} & \textbf{0.87} \\
\texttt{WBTC-WETH} 0.05\%   & \textbf{-2.27} & \textbf{0.25} & \textbf{0.82} & \textbf{-2.48} & \textbf{0.13} & \textbf{0.66} & \textbf{-2.04} & \textbf{0.15} & \textbf{0.94} \\
\texttt{WBTC-WETH} 0.3\%    & \textbf{-0.72} & \textbf{0.36} & \textbf{0.91} & \textbf{-2.26} & \textbf{0.31} & \textbf{0.52} & 0.34 & 0.69 & 0.42 \\
\texttt{LINK-WETH} 0.3\%    & \textbf{1.37}  & \textbf{0.81} & \textbf{0.78} & \textbf{-2.39} & \textbf{0.12} & \textbf{0.64} & -2.73 & 0.50 & 0.22 \\
\texttt{MNT-WETH} 0.3\% & -0.13 & 0.63 & 0.45 & 0.01 & 0.66 & 0.40 & \textbf{-2.53} & \textbf{0.13} & \textbf{0.39} \\
\texttt{UNI-WETH} 0.3\% & \textbf{0.02}  & \textbf{0.47} & \textbf{0.77} & \textbf{-1.94} & \textbf{0.16} & \textbf{0.83} & -5.28 & 0.08 & 0.01 \\

\midrule

\texttt{USDC-USDT} 0.01\%   & \textbf{-4.82} & \textbf{0.28} & \textbf{0.48} & \textbf{-2.96} & \textbf{0.29} & \textbf{0.56} & \textbf{-2.39} & \textbf{0.41} & \textbf{0.75} \\
\texttt{USDC-USDT} 0.05\%   & -7.40 & 0.10 & 0.00 & \textbf{-2.06} & \textbf{0.10} & \textbf{0.56} & \textbf{-1.62} & \textbf{0.13} & \textbf{0.70} \\
\texttt{DAI-USDC} 0.01\%    & 0.65 & 1.37 & 0.55 & -3.94 & 0.21 & 0.11 & \textbf{-1.37} & \textbf{0.47} & \textbf{0.73} \\
\texttt{DAI-USDT} 0.01\%    & -3.09 & 0.85 & 0.24 & -1.43 & 0.49 & 0.37 & -2.90 & 0.46 & 0.20 \\
\texttt{USDe-USDT} 0.01\%   & -3.57 & 0.87 & 0.12 & \textbf{-1.90} & \textbf{0.10} & \textbf{0.81} & \textbf{-1.66} & \textbf{0.04} & \textbf{0.81} \\

\midrule

\texttt{WETH-weETH} 0.01\%  & 0.05 & 0.98 & 0.19 & -0.36 & 0.78 & 0.37 & \textbf{0.37}  & \textbf{0.63} & \textbf{0.66} \\
\texttt{wstETH-WETH} 0.01\% & -9.84 & -0.20 & 0.02 & \textbf{-3.20} & \textbf{0.20} & \textbf{0.27} & \textbf{-0.44} & \textbf{0.58} & \textbf{0.77} \\
\texttt{WBTC-LBTC} 0.05\%   & 9.60 & 3.69 & 0.63 & -3.43 & 0.33 & 0.09 & \textbf{-0.00} & \textbf{0.34} & \textbf{0.75} \\
\bottomrule
\end{tabular}
\caption{Power-law fit results from the regression $ACF(L) \sim A \cdot L^{-p}$. \textbf{R2} is for the R2 score of the linear regression $\log(ACF(L)) \sim \log(A) - p\log(L)$. The left, central, and right columns are for the returns' absolute value, traded volume, and swap signs, respectively. Bold cells indicate long-memory.}
\label{tab:lm_rrts}
\end{table}

First, we observe that the pool fee tier could change the statistical properties of the pool. That is, fixed one pair, if there is ACF persistence in the 0.05\% pool, it is not certain to recover the same pattern in the 0.01\% or 0.3\% pools. As for Normal pairs, we notice that: pools involving USDC or USDT similarly behave (that is, there is long-memory for the same ACFs and fee tiers); and the exponent $p$ typically increases with the pool fee tier. Regarding short-memory in the volume time series ACF, we notice just a few occurrences, which are overlapping with deviations in the transition probabilities patterns and with the least active pools -to have a look, compare Table \ref{tab:trans_prob} and central columns in Table \ref{tab:lm_rrts}. As for long-memory in the swap signs series, we find this pattern to be more frequently shown in Synthetic and Stable pools. These also show significantly bigger exponent $p$ and coefficient $\log(A)$.\\

Finally, we focus on the relationship between volume and Realized Variance (RV), which is used as a proxy for the integrated variance. In Section \ref{subsec:vol_sign} of the main text, we show that several pools exhibit a significant linear relationship between the logarithms of volume and RV. We fit the linear regression $\log(Vol) = A + q\log(RV)$ to verify this. Specifically, the Volume is intended to be the aggregated volume, and both $Vol$ and $RV$ are computed over a time window of $n$ minutes, with $n=1,5, 10, 30$. Table \ref{tab:vol_vol} shows the fitted coefficient $q$ and the $R2$ score. The results show $q\approx 0.5$. There is not a monotonic relationship of $q$ from $n$, as it seems to be strongly pool-dependent.
\begin{table}[h]
\centering
\begin{tabular}{lrr|rr|rr|rr}
\toprule
\textbf{Pool}    & \multicolumn{2}{c|}{\textbf{1 min}} & \multicolumn{2}{c|}{\textbf{5 min}} & \multicolumn{2}{c|}{\textbf{10 min}} & \multicolumn{2}{c}{\textbf{30 min}} \\
\multicolumn{1}{r}{\textbf{}} & \textbf{q} & \textbf{R2} & \textbf{q} & \textbf{R2} & \textbf{q} & \textbf{R2} & \textbf{q} & \textbf{R2} \\
\midrule
\midrule
\texttt{USDC-WETH} 0.01\% & \multicolumn{1}{l}{0.46} & \multicolumn{1}{l|}{0.38} & \multicolumn{1}{l}{0.54} & \multicolumn{1}{l|}{0.33} & \multicolumn{1}{l}{0.67} & \multicolumn{1}{l|}{0.37} & \multicolumn{1}{l}{0.95} & \multicolumn{1}{l}{0.51} \\
\texttt{USDC-WETH} 0.05\% & \textbf{0.47}& \textbf{0.91} & \underline{ 0.44} & \underline{ 0.80} & 0.43 & 0.74 & 0.42 & 0.68 \\
\texttt{USDC-WETH} 0.3\% & \textbf{0.49}& \textbf{0.92} & \textbf{0.50}& \textbf{0.93} & \textbf{0.51}& \textbf{0.93} & \textbf{0.53}& \textbf{0.93}\\
\texttt{WETH-USDT} 0.01\% & \underline{ 0.49} & \underline{ 0.76} & 0.45 & 0.51 & 0.42 & 0.37 & 0.49 & 0.25 \\
\texttt{WETH-USDT} 0.05\% & \underline{ 0.46} & \underline{ 0.84} & 0.44 & 0.64 & 0.41 & 0.53 & 0.37 & 0.43 \\
\texttt{WETH-USDT} 0.3\% & \textbf{0.49}& \textbf{0.95} & \textbf{0.50}& \textbf{0.95} & \textbf{0.51}& \textbf{0.95} & \textbf{0.52}& \textbf{0.93}\\
\texttt{WBTC-USDC} 0.05\% & \multicolumn{1}{l}{0.40} & \multicolumn{1}{l|}{0.46} & \multicolumn{1}{l}{0.44} & \multicolumn{1}{l|}{0.49} & \multicolumn{1}{l}{0.46} & \multicolumn{1}{l|}{0.50} & \multicolumn{1}{l}{0.49} & \multicolumn{1}{l}{0.48} \\
\texttt{WBTC-USDC} 0.3\% & 0.38 & 0.54 & 0.41 & 0.57 & 0.43 & 0.59 & 0.46 & 0.61 \\
\texttt{WBTC-USDT} 0.05\% & \multicolumn{1}{l}{0.27} & \multicolumn{1}{l|}{0.14} & \multicolumn{1}{l}{0.28} & \multicolumn{1}{l|}{0.12} & \multicolumn{1}{l}{0.30} & \multicolumn{1}{l|}{0.11} & \multicolumn{1}{l}{0.36} & \multicolumn{1}{l}{0.11} \\
\texttt{WBTC-USDT} 0.3\% & 0.44 & 0.70 & 0.46 & 0.72 & 0.47 & 0.73 & 0.5 & 0.73 \\
\midrule
\texttt{WBTC-WETH} 0.01\% & 0.35 & 0.40 & 0.31 & 0.16 & 0.32 & 0.12 & 0.37 & 0.11 \\
\texttt{WBTC-WETH} 0.05\% & \underline{ 0.46} & \underline{ 0.88} & \underline{ 0.46} & \underline{ 0.85} & \underline{ 0.46} & \underline{ 0.82} & 0.44 & 0.68 \\
\texttt{WBTC-WETH} 0.3\% & \underline{ 0.45} & \underline{ 0.88} & \textbf{0.46}& \textbf{0.90} & \textbf{0.47}& \textbf{0.90} & \underline{ 0.48} & \underline{ 0.89} \\
\texttt{LINK-WETH} 0.3\% & \underline{ 0.46} & \underline{ 0.85} & \underline{ 0.48} & \underline{ 0.86} & \underline{ 0.49} & \underline{ 0.86} & \underline{ 0.50} & \underline{ 0.84} \\
\texttt{MNT-WETH} 0.3\%& \textbf{0.50}& \textbf{1.00} & \textbf{0.51}& \textbf{1.00} & \textbf{0.51}& \textbf{0.99} & \textbf{0.52}& \textbf{0.99}\\
\texttt{UNI-WETH} 0.3\%& \textbf{0.48}& \textbf{0.93} & \textbf{0.50}& \textbf{0.94} & \textbf{0.50}& \textbf{0.93} & \textbf{0.52}& \textbf{0.92}\\
\midrule
\texttt{USDC-USDT} 0.01\% & 0.36 & 0.74 & 0.35 & 0.71 & 0.33 & 0.66 & 0.25 & 0.52 \\
\texttt{USDC-USDT} 0.05\% & \multicolumn{1}{l}{\underline{ 0.42}} & \multicolumn{1}{l|}{\underline{ 0.86}} & \multicolumn{1}{l}{\underline{ 0.43}} & \multicolumn{1}{l|}{\underline{ 0.85}} & \multicolumn{1}{l}{\underline{ 0.44}} & \multicolumn{1}{l|}{\underline{ 0.84}} & \multicolumn{1}{l}{\underline{ 0.45}} & \multicolumn{1}{l}{\underline{ 0.84}} \\
\texttt{DAI-USDC} 0.01\% & \textbf{0.47}& \textbf{0.93} & \textbf{0.48}& \textbf{0.95} & \textbf{0.49}& \textbf{0.95} & \textbf{0.50}& \textbf{0.94}\\
\texttt{DAI-USDT} 0.01\% & \multicolumn{1}{l}{\underline{ 0.43}} & \multicolumn{1}{l|}{\underline{ 0.85}} & \multicolumn{1}{l}{\underline{ 0.43}} & \multicolumn{1}{l|}{\underline{ 0.84}} & \multicolumn{1}{l}{\underline{ 0.43}} & \multicolumn{1}{l|}{\underline{ 0.83}} & \multicolumn{1}{l}{\underline{ 0.42}} & \multicolumn{1}{l}{\underline{ 0.79}} \\
\texttt{USDe-USDT} 0.01\% & 0.40 & 0.73 & \underline{ 0.40} & \underline{ 0.76} & \underline{ 0.40} & \underline{ 0.75} & 0.37 & 0.66 \\
\midrule
\texttt{WETH-weETH} 0.01\% & \underline{ 0.41} & \underline{ 0.79} & \underline{ 0.41} & \underline{ 0.76} & 0.40 & 0.72 & 0.36 & 0.57 \\
\texttt{wstETH-WETH} 0.01\% & \underline{ 0.40} & \underline{ 0.80} & \underline{ 0.41} & \underline{ 0.81} & \underline{ 0.40} & \underline{ 0.80} & \underline{ 0.37} & \underline{ 0.75} \\
\texttt{WBTC-LBTC} 0.05\% & \underline{ 0.44} & \underline{ 0.84} & \underline{ 0.42} & \underline{ 0.78} & 0.40 & 0.73 & 0.37 & 0.61  \\
\bottomrule
\end{tabular}
\caption{Volume vs Realized Variance - fitted coefficient $q$ and $R2$ score of the linear regression $\log(Vol) = \log(A) + q\log(RV)$. The fits with $R2\ge0.90$ are highlighted in bold; fits with $R2\ge0.75$ are underlined.}
\label{tab:vol_vol}
\end{table}

\end{document}